\documentclass[aps,prb,reprint,groupedaddress,floatfix]{revtex4-1}

\usepackage[english]{babel} 
\usepackage{amssymb,amsmath}
\usepackage{graphicx}
\usepackage{url}
\usepackage{multirow}
\usepackage{chemformula}
\usepackage{tabularx}
\usepackage{makecell}
\usepackage{xcolor}
\usepackage{siunitx}

\usepackage{subcaption}

\newlength{\smallpic}
\begin{document}

\title{\emph{Ab initio} property characterisation of thousands of previously unexplored 2D materials}
\author{Peder Lyngby$^*$ and Kristian Sommer Thygesen
 }
 \address{Computational Atomic-scale Materials Design (CAMD), Department of Physics, Technical University of Denmark, 2800 Kgs. Lyngby Denmark \\ \normalfont{$^*$Corresponding author: pmely@dtu.dk}}
\date{\today}

\begin{abstract}
We perform extensive density functional theory (DFT) calculations to determine the stability and elementary properties of 4249 previously unexplored monolayer crystals. The monolayers comprise the most stable subset (energy within 0.1 eV/atom of the convex hull) of a larger portfolio of two-dimensional (2D) materials recently discovered using a deep generative model and systematic lattice decoration schemes. The relaxed 2D structures are run through the basic property workflow of the Computational 2D Materials Database (C2DB) to evaluate the dynamical stability and obtain the stiffness tensor, piezoelectric tensor, deformation potentials, Born and Bader charges, electronic band structure, effective masses, plasma frequency, Fermi surface, projected density of states, magnetic moments, magnetic exchange couplings, magnetic anisotropy, topological indices, optical- and infrared polarisability. We provide statistical overviews of the property data and highlight a few specific examples of interesting materials. Our work exposes previously unknown parts of the 2D chemical space and provides a basis for the discovery of 2D materials with specific properties. All data is available in the C2DB.    
\end{abstract}

\maketitle

\section{Introduction}
The combination of powerful \emph{ab initio} modeling codes and artificial intelligence (AI) models is opening new opportunities in materials science. Today, it is possible to determine the atomic and electronic structure of even fairly complex materials within a few hours on a single compute node implying that thousands of materials can be scrutinised in a time span of weeks on standard high-performance computing clusters. Such data can then be curated and organised in databases\cite{marzari2021electronic,thygesen2016making,saal2013materials,curtarolo2012aflow} on which AI models can be trained to establish structure-property relations\cite{montavon2013machine,ward2016general,manti2023exploring}. Over the past few years numerous examples of such applications have been demonstrated, and AI has already become an integral part of the computational materials scientist's toolbox when it comes to modeling of structures and predicting their properties\cite{wei2019machine,mueller2016machine,himanen2019data,schmidt2019recent,knosgaard2022representing}. Compared to such supervised learning tasks, a far more intricate and challenging application of AI is the unsupervised generation of new materials (composition and structure) with prescribed properties -- the most basic such property being the thermodynamic stability of the material.

Within the past couple of years, computational materials generation projects (some of them partly AI-driven) have led to the discovery of hundreds of thousands of previously unknown inorganic bulk crystals with high thermodynamic stability, i.e. at or very close to the so-called convex hull\cite{merchant2023novel, schmidt2022a,schmidt2022largescale}. It must be expected that many of these compounds can be synthesised, and thus they represent an enormous reservoir of candidate materials some of which could be used to improve the performance of existing technologies or even enable new ones. The in-silico expansion of the set of known inorganic crystals by almost an order of magnitude within a couple of years is a tremendous intellectual achievement. However, for practical purposes knowledge of the structure and composition of stable materials is not very useful in itself because the decision to synthesise and deploy a given material usually requires some presumptions about the material's properties. Therefore, to make 
computational materials discovery relevant for experiments and applications, the determination of stable crystal structures must be complemented by a characterisation of the most basic materials properties.

Two-dimensional (2D) materials represent an emerging class of materials whose unique and unconventional properties make them interesting for both fundamental science and technological applications. There exist around 800 layered bulk materials in experimental crystal structure databases whose layer-layer bonds are predicted by \emph{ab initio} calculations to be sufficiently weak that single layers can be exfoliated\cite{ashton2017topology,mounet2018two}. The vast majority of these materials, including the around 100 that have already been produced in mono- or few-layer form\cite{haastrup2018computational}, are contained in the Computational 2D Materials Database (C2DB) since 2021\cite{gjerding2021atomic}. 

Recently, we trained a crystal diffusion variational autoencoder (CDVAE) \cite{xie2021crystal} on the most stable crystal structures in the C2DB and used it to generate a large set of new, thermodynamically stable 2D materials\cite{lyngby2022data}. The set of CDVAE-generated crystals was complemented by 2D crystals generated by a more traditional lattice decoration approach where the atoms in the known structures were substituted by chemically similar ones. After removing duplicates and non-2D crystal structures, this resulted in a portfolio of 11.630 previously unknown 2D crystals, which were subsequently relaxed using density functional theory (DFT) calculations. 
Concurrently with our work, Wang \textit{et al} \cite{wang2023symmetry} discovered around 6500 low-energy 2D materials by systematically occupying all the Wyckoff positions of selected layered space groups by all possible atoms. 

In the present work, we calculate the elementary physical properties of the most stable subset of monolayers resulting from our own structure generation project\cite{lyngby2022data} as well those found by Wang \textit{et al} \cite{wang2023symmetry}. Specifically, we consider all monolayers with an energy above the convex hull of less than 100 meV/atom. This amounts to a total of 4249 materials of which 629 originate from the CDVAE, 2702 were produced by lattice decoration while the remaining 918 come from Wang \textit{et al}. As a testimony to the good thermodynamic stability and chemical validity of the structures, we find that the majority of them (about 70\%) are dynamically stable, i.e. stable against small perturbations of the atom positions and unit cell shape. For this subset of 2759 materials we employ the computational workflow behind the C2DB to compute a wide variety of properties, leaving the more computationally demanding steps of the workflow (e.g. GW quasiparticle band structures, absorbance spectra including excitonic effects, and nonlinear optical properties) for future work. All the structures and properties are available on on the C2DB website where they can be browsed or downloaded.

Prior to the present work, the C2DB (which is currently the largest 2D materials database) contained 1345 monolayers with convex hull energy below 0.1 eV/atom. Thus the new set of monolayers characterised in this work, triples the number of (theoretically) known stable 2D materials.

\begin{table*}[!ht]
   
    {\renewcommand{\arraystretch}{1.5}
    \begin{tabular}{lllr}

        Property & Method & Criteria & Count  \\ \hline\hline

        Energy above convex hull & PBE & None &  4249\\ \hline

        Heat of formation & PBE & None &  4249 \\ \hline

        Electronic band structure PBE & PBE* & None &  4139 \\ \hline

        Orbital projected band structure & PBE & None  &  4139 \\ \hline

        Out-of-plane dipole & PBE & None &  4139 \\ \hline

        Work function & PBE* & None &  4139 \\ \hline

        Bader charges & PBE & None & 3525\\ \hline

        Projected density of states & PBE & None &  4139 \\ \hline

        Phonons ($\Gamma$ and BZ corners) & PBE & None &  3923 \\ \hline

        Stiffness tensor & PBE & None &  4085 \\ \hline

        Exchange couplings & PBE & DS, Magnetic,  $N_{\mathrm{atoms}}^{\mathrm{magn.}} \le 2$ &  196 \\ \hline

        Magnetic anisotropies & PBE* & DS, Magnetic &  744 \\ \hline

        Deformation potentials & PBE* & DS, $E^\mathrm{PBE}_{\mathrm{g}}>0$ &  1142\\ \hline

        Effective masses & PBE* & DS, $E^\mathrm{PBE}_{\mathrm{g}}>0$ &  1969 \\ \hline

        Fermi surface & PBE* & DS, $E^\mathrm{PBE}_{\mathrm{g}}=0$ &  763 \\ \hline

        Plasma frequency & PBE* & DS, $E^\mathrm{PBE}_{\mathrm{g}}=0$, $N_{\mathrm{atoms}}<15$ &  628 \\ \hline

        Infrared polarisability & PBE & DS, $E^\mathrm{PBE}_{\mathrm{g}}>0$, $N_{\mathrm{atoms}}<15$ &  1090 \\ \hline

        Optical polarisability & RPA@PBE & DS, $N_{\mathrm{atoms}}<15$  &  1833 \\ \hline

        Electronic band structure & HSE06@PBE* & DS, $E^\mathrm{PBE}_{\mathrm{g}}>0$ &  1922 \\ \hline

        Born charges & PBE, Berry phase & DS, $E^\mathrm{PBE}_{\mathrm{g}}>0$&  1826 \\ \hline

        Piezoelectric tensor & PBE, Berry phase & DS, $E^\mathrm{PBE}_{\mathrm{g}}>0$, non-centrosym. &  858 \\ \hline

        Topological invariants & PBE*, Berry phase & DS, $0<E^\mathrm{PBE}_{\mathrm{g}}<0.7$ eV, $N_{\mathrm{atoms}}<13$ & 340 \\ \hline

        Half-metal gap & PBE & DS, $E^\mathrm{PBE}_{\mathrm{g}}=0$, Magnetic & 254 \\ \hline
        
        Half-metal gap & HSE06@PBE & DS, $E^\mathrm{PBE}_{\mathrm{g, half}}>0$, Magnetic & 74 

    \end{tabular}}

    \caption{Properties calculated by the C2DB workflow. The computational method and the criteria used to decide whether the property should be evaluated for a given material is also shown. The symbol '*' indicates that spin--orbit coupling (SOC) is included. All calculations are performed with the GPAW electronic structure code using a plane wave basis set. The variations in count numbers among properties sharing identical criteria stem from convergence problems in the DFT calculations. DS: Dynamical stable. $E_{\mathrm{g, half}}$: Half-metal gap.}

    \label{tab:properties}

\end{table*}

\begin{figure*}
    \centering
    \includegraphics[]{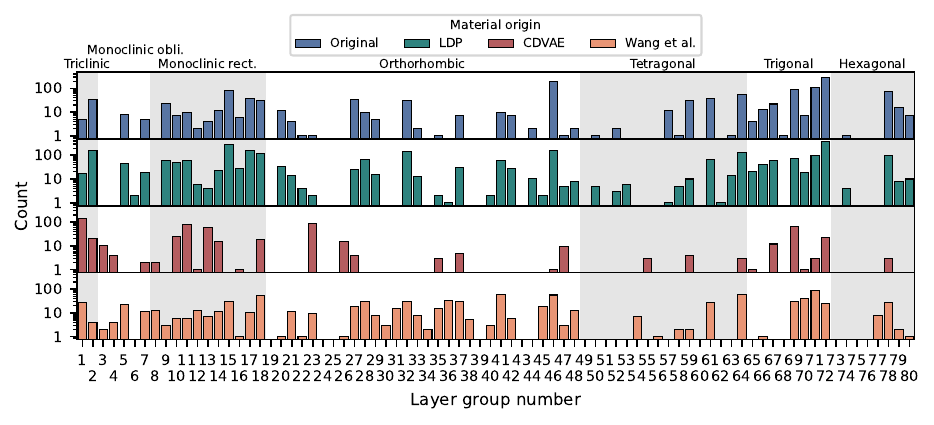}
    \caption{Histograms showing the distribution of the 2D crystals investigated in this work according to their layer group number. Note the logarithmic scale. Histograms are shown separately for the four distinct groups of materials: The original stable materials of the C2DB (original), the materials generated by the deep generative crystal diffusion model (CDVAE), the materials generated by lattice decoration of the original C2DB materials (LDP), and the materials generated by the symmetry-based approach of Wang \emph{et al.} (Wang et al.). The layer group has been determined with the algorithm in Ref. \cite{fu2024layer} using a symmetry precision threshold of 0.1 \AA. }
    \label{fig:layergroup}
\end{figure*}

\begin{figure*}
    \centering
    \includegraphics[]{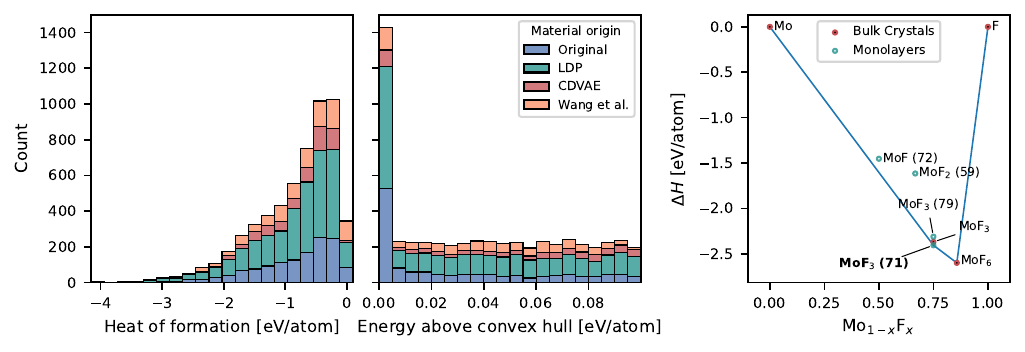}
    \caption{Stacked histograms showing the distribution of heat of formation, $\Delta H$, and energy above the convex hull, $\Delta H_{\mathrm{hull}}$, for the different materials investigated in this work. To the right: Example of the convex hull for the LDP generated material of MoF$_3$ (highlighted in bold), which lies on the convex hull. The numbers in brackets refers to the layer group number of the monolayer. Monolayers (bulk crystals) are indicated by an open (filled) circle. The bulk crystals used to generate the convex hull are the most stable compounds of he Open Quantum Materials Database (OQMD) with up to three different atoms types.}
    \label{fig:ehull}
\end{figure*}

\section{Results}
In this section we first introduce the 2D materials investigated in this work and describe the different types of properties calculated. A few selected materials with specific or particularly interesting properties are highlighted along the way for illustrative purposes. Out of the 4249 materials with formation energies within 0.1 eV/atom of the convex hull, 2759 are found to be dynamically stable. Further property calculations are performed for this subset only. A complete list of the properties explored and the number of materials for which each property has been determined, is provided in Table \ref{tab:properties}. The table does not comprise all properties of the C2DB workflow. The remaining properties will be computed at a later stage.

For consistency, all calculations presented in this work employ the xc-functionals PBE and HSE06 (only for band structures); see Method section for more details. It is well known that materials with highly localised, partially filled states (e.g open $d$-shells) can exhibit strong correlation effects, which are poorly described by PBE (and HSE06). For this reason, we have also performed systematic PBE+U calculations for all the materials containing one or more of the 3d transition metals V, Cr, Mn, Fe, Co, Ni. An analysis of these results will be presented elsewhere and all the results will be available in C2DB.

\subsection{Generation of materials}
Recently, we used a crystal diffusion variational autoencoder (CDVAE) to generate 5003 monolayer structures\cite{lyngby2022data}. The CDVAE model was trained on a set of 2615 monolayers from the Computational 2D Materials Database (C2DB)\cite{haastrup2018computational,gjerding2021recent} with energy above the convex hull, $\Delta H_{\mathrm{hull}}$, below $0.3 \; \mathrm{eV/atom}$. We shall refer to this set of materials as the "seed structures". The CDVAE-generated materials were complemented by 14192 monolayers obtained by replacing atoms in the seed structures by chemically similar ones. We shall refer to this procedure as the lattice decoration protocol (LDP). The generated structures were subsequently relaxed using DFT and their heat of formation and energy above the convex hull calculated. In the end 3073 (8579) unique 2D structures were obtained using the CDVAE (LDP) generation models.

Importantly, we found that both the CDVAE and LDP generated crystals inherited the good stability properties of the seed structures. As a side remark we mention that the CDVAE- generated structures showed high degree of structural and chemical diversity extending beyond that of the seed structures and the lattice decorated structures. For more details on the CDVAE and LDP generation processes, along with a comprehensive comparison we refer to Ref. \cite{lyngby2022data}. 

For the present work we have selected the 3331 most stable subset of the 2D materials generated by the CDVAE and LDP, namely those with $\Delta H_{\mathrm{hull}}< 0.1$ eV/atom. 

The materials from Wang \textit{et al} \cite{wang2023symmetry} are all binary and ternary compounds and are primarily generated by a symmetry-based approach where the different Wyckoff positions of a given layer group are occupied by up to three different types of elements. This approach is free of any structural or chemical bias originating from a set of seed structures (as is the case for the LDP and to some extent the CDVAE). However, this symmetry-based generation of materials does also not contain any bias towards producing stable materials (as is the case for both LDP and CDVAE). Consequently, Wang \textit{et al.} invoke several screening criteria to remove unstable materials before the final DFT relaxation. These include conditions on charge neutrality and electronegativity as well as a pre-relaxation of the structure using a machine learning universal interatomic potential. Additionally, Wang \textit{et al.} performed atom substitution with all elements from the periodic table for a subset of the generated materials and used a crystal-graph neural network to screen the resulting structures for stability. 
All the structures generated by Wang \textit{et al.} with $\Delta H_{\mathrm{hull}}< 0.1$ eV/atom have been re-relaxed using the same computational framework as used for the CDVAE and LDP structures (see Methods). In addition, we have removed any structures categorized as non-2D by our dimensionality analysis\cite{gjerding2021recent} and 81 structures which were already present in the C2DB or the set of CDVAE/LDP-generated materials. 

The combined set of new materials (presenting no overlap with the original structures in C2DB) comprises 4249 unique crystals with $\Delta H_{\mathrm{hull}}< 0.1$ eV/atom. This set of crystals contains 3610 unique (reduced) chemical formulas of which only 356 are present in the original C2DB. In the following we explore the properties of the 4249 new materials and compare them to the subset of structures in the C2DB with $\Delta H_{\mathrm{hull}}< 0.1$ eV/atom, referred to as the set of stable original materials.

\subsection{Crystal symmetry}
The layer group (the analogue of the space group for a crystal with a 2D lattice) has been determined for all the materials using the algorithm in Ref. \cite{fu2024layer}. In Fig. \ref{fig:layergroup} we show the histograms of layer group numbers for each of the four sets of materials, namely the original stable materials of the C2DB (original), the materials generated by the deep generative crystal diffusion model (CDVAE), the materials generated by the lattice decoration protocol (LDP), and the materials generated by the symmetry-based approach of Wang \emph{et al.} (Wang et al.). Note the logarithmic scale. Only materials with formation energies within 0.1 eV/atom of the convex hull are included in the plot. 

Not surprisingly, the layer group distribution of the LDP materials is very similar to that of the original materials from which they were generated. The CDVAE structures contain relatively few layer groups while the structures of Wang \emph{et al.} are more homogeneously distributed and span a larger set of layer groups. No materials are present in the data set for 11 out of the 80 layer groups. These layer groups are: 19, 24, 25, 39, 43, 49, 51, 60, 73, 75, and 76.

\subsection{Thermodynamic stability}
All the monolayers considered have formation energies within 0.1 eV/atom of the convex hull. In this work we use a convex hull defined by a reference database consisting of the most stable 1,2, and 3-element solid phases from the Open Quantum Materials Database (OQMD)\cite{kirklin2013high} amended by the monolayers themselves (this ensures that the energy above the hull is always non-negative). The threshold value of 0.1 eV/atom has been chosen to 
account for the finite accuracy of the PBE xc-functional, and to allow for inclusion of meta-stable crystal structures. There are dozens of examples of meta-stable 2D crystal structures that have been experimentally characterised, e.g. the T'-phases of Mo-based dichalcogenides with $\Delta H_{\mathrm{hull}}$ ranging from 0.18 to 0.02 eV/atom.

The distribution of heat of formation, $\Delta H$, and energy above the convex hull, $\Delta H_{\mathrm{hull}}$, for the new materials and stable original materials is shown in Fig. \ref{fig:ehull}. An example of an LDP-generated monolayer on the convex hull ($\Delta H_{\mathrm{hull}}=0$ eV/atom) is MoF$_3$ as seen on the right panel. This monolayer phase has a formation energy of $0.04$ eV/atom below the most stable bulk phase of the same composition. We note that MoF$_3$ (like all the monolayers studied here) does not have a known layered bulk counterpart and thus could not have been found by analysing bulk crystal structure databases.

In total 735 monolayers are predicted to lie exactly on the convex hull. We stress that in reality a monolayer would never lie exactly on the convex hull because its layered bulk form would always have a lower energy due to the attractive interlayer forces. Such layered bulk crystals are not included in the reference database of bulk crystals used to calculate the convex hull, and therefore the monolayers can end up be exactly on-hull. 
For monolayers with high thermodynamic stability, like the ones studied here, interlayer interactions will be weak and mostly of the van der Waals type (see next section). Therefore, we can safely assume that the monolayers will remain close to the convex hull even if the layered bulk form was included in the convex hull. On a more technical note, a correct description of the interlayer interactions in vdW bulk crystals require the use of an xc-functional that can account for dispersive forces. In our work we use the PBE functional that does not capture such forces. Thus, even if the layered bulk form of a given monolayer would be included in the convex hull, it would only be predicted to have a lower energy than the monolayer if the layers form stronger chemical (covalent/ionic) bonds.

\subsection{Exfoliability}
In general, there are two main routes to the production of monolayer crystals, namely chemical synthesis (e.g. via atomic layer deposition, chemical vapor deposition, or pulsed laser deposition) or exfoliation from a bulk crystal (either mechanical or via selective chemical etching). The two approaches may be viewed as bottom-up or top-down, respectively.

Common to all the monolayers studied in this work is that they do not have a known (naturally occurring or previously synthesised) layered bulk analogue. Consequently, they would have to be synthesised bottom-up. Here it should be noted that chemical synthesis might result in a film containing several monolayers. In such cases, an exfoliation step is required to isolate a single monolayer. 

An important descriptor for evaluating the synthesisability of a monolayer crystal is its exfoliation energy, i.e. the difference in total energy between the isolated monolayer and the layered bulk (per layer) normalised by the surface area. When the parent bulk structure is known, the calculation of the exfoliation energy is in principle straightforward\cite{mounet2018two}. When this is not the case, the calculation is more involved as various stacking configurations must be explored in order to find the most stable layered structure\cite{pakdel2023emergent}. Based on an analysis of the exfoliation energies calculated using DFT-vdW for a set of known materials, Mounet \emph{et al.}\cite{mounet2018two} concluded that an exfoliation energy of 30 meV/\AA$^2$ is a reasonable upper limit for "easily exfoliable" materials, and that materials with exfoliation energies up to 150 meV/\AA$^2$ are "potentially exfoliable".

The size of the exfoliation energy is largely determined by the chemical inertness of the monolayer (e.g. absence of dangling bonds). Thus a high thermodynamic stability of the monolayer should be indicative of low exfoliation energy. This expectation is supported by our recent work\cite{pakdel2023emergent} where we perform DFT-vdW calculations of the exfoliation energies of 1052 monolayers in the original C2DB dataset with $\Delta H_{\mathrm{hull}}=0.2$ eV/atom and found that the vast majority of the monolayers had exfoliation energies below 50 meV/\AA$^2$. Based on these findings it is very likely that the monolayers of the present study will have comparably low exfoliation energies.

\begin{figure*}
    \centering
    \includegraphics[]{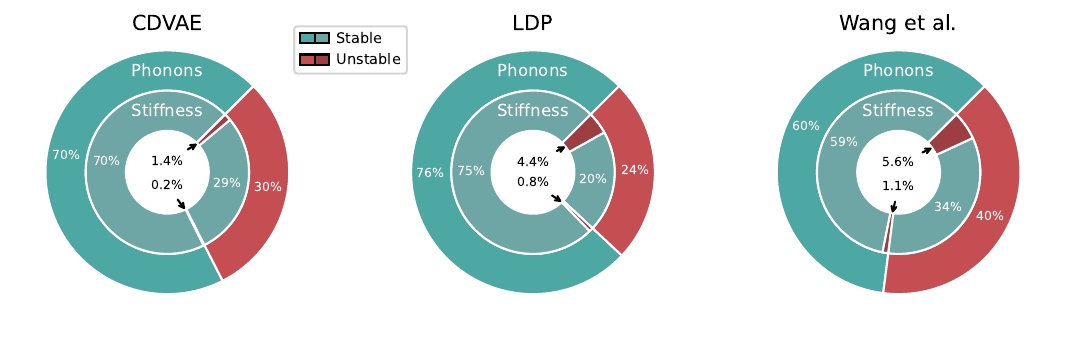}
    \caption{Overview of the dynamical stability of the new materials for the three material groups. Green (red) colors denote stable (unstable) materials. The test for dynamical stability involves a separate test of the phonon frequencies and the eigenvalues of the stiffness tensor. The diagrams show the result of both tests (inner and outer rims). It can be noted that materials are much more likely to be dynamically unstable due to phonons (distortions of the atomic structure) than the stiffness tensor (instability of the unit cell shape). }
    \label{fig:dynstab}
\end{figure*}

\begin{figure*}
    \centering
    \includegraphics[]{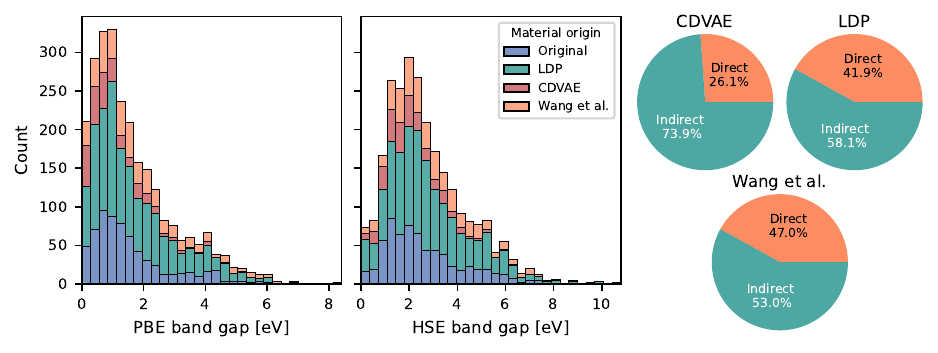}
    \caption{Left: Stacked histograms showing the distribution of band gaps (metals not included) for the newly generated materials and the stable original materials in the C2DB. Band gaps obtained using the PBE and HSE06 xc-functionals are shown in the left and right histograms, respectively. Right: The pie charts show the fraction of the newly generated materials with a direct and indirect PBE band gap, respectively. All band energies include spin-orbit coupling.}
    \label{fig:gap_hist}
\end{figure*}

\subsection{Dynamical stability}
The dynamical stability is assessed by calculating the phonons at the high-symmetry points of the Brillouin zone corresponding to the $q$-points (0,0), (0,0.5), (0.5,0), and (0.5,0.5) in fractional coordinates. An imaginary frequency signals a phonon instability. Moreover, the stiffness tensor is calculated and diagonalized. A negative eigenvalue signals an instability of the shape of the unit cell. A material is termed dynamically stable if all phonon frequencies and stiffness tensor eigenvalues are real and positive. More details including justification of the scheme can be found in Ref. \cite{manti2023exploring}.

Figure \ref{fig:dynstab} shows the distribution of materials according to the dynamical stability (phonons and stiffness, respectively). The materials have been subdivided according to their origin, i.e. LDP, CDVAE, and Wang \emph{et al.}. For all groups of materials, phonon instabilities occur more frequently than stiffness instabilities. In particular, very few materials that are stable with respect to phonons show stiffness instability (0.2\%, 0.8\%, and 1.1\% for the three groups of materials). A slightly higher percentage (76\%) of the LDP materials are found to be dynamically stable with respect to phonons as compared to the CDVAE materials (70\%), which in turn have a higher phonon stability rate than the   materials from Wang \emph{et al.} (60\%).

The percentage of dynamically stable structures produced by the LDP and CDVAE schemes (70-76\%) is similar to that of the subset of seed structures with $\Delta H_{\mathrm{hull}}<0.1$ eV/atom (76\%) suggesting that the LDP/CDVAE models learn to generate dynamically stable crystals from the seed/training structures.  

In total 2759 of the new materials are found to be dynamically stable. The remaining part of the property workflow is limited to this set of materials.

\subsection{Electronic band structure}

\begin{figure*}
    \centering
    \includegraphics[]{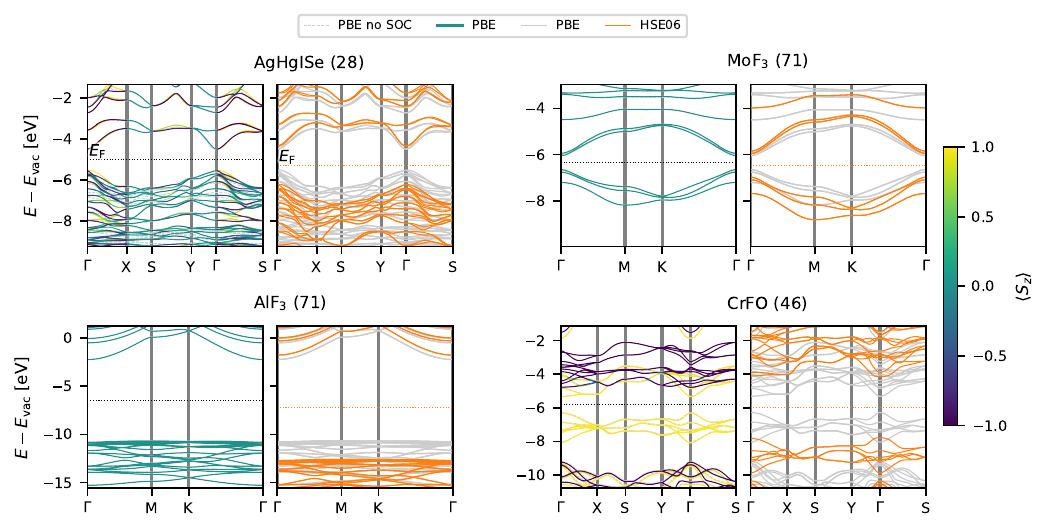}
    \caption{Examples of band structures. The band structure calculated using the PBE xc-functional with and without spin orbit coupling as well as HSE06 hybrid functional is shown for the four materials AgHgISe, MoF$_3$, AlF$_3$, and CrFO. The four selected materials have direct band gaps of varying sizes, are dynamically stable, and have energy above the convex hull of at most $0.03$ eV/atom (MoF$_3$ and CrFO lie on the convex hull). CrFO has a ferromagnetic ground state.}
    \label{fig:BS}
\end{figure*}

The electronic band structure has been calculated for all the dynamically stable materials using the PBE xc-functional. For materials with a finite band gap, the band structure has also been obtained with the HSE06 hybrid functional. In both cases spin-orbit coupling (SOC) is included.

The distribution of the PBE and HSE06 band gaps is shown in Fig. \ref{fig:gap_hist} (metals not included). About 30\% the new materials generated by LDP and CDVAE are metallic, which is very similar to that of the seed structures (31 \%). The fraction of metallic compounds is significantly lower in the set from Wang \emph{et al.} (11\%). Focusing on the 1982 non-metallic compounds, we find that 26\% (CDVAE), 42\% (LDP), and 47\% (Wang) of the new structures have a direct band gap, see pie charts on Fig. \ref{fig:gap_hist}. Here we classify band gaps as direct if the direct gap is below or within 5 meV of the indirect gap.  

Ultra wide band gap materials are important for applications such as insulating dielectrics and ultraviolet photonics. We find 18 new 2D materials with very large band gaps, here defined as the HSE06 band gap exceeding 8 eV. In comparison, the largest band gap in the original set of stable materials in C2DB is 7.49 eV (MgB$_2$H$_8$). 17 of the new large band gap materials are fluorides and most of them have chemical formula XYF$_6$ or YF$_3$, where X is an alkali metal and Y is a transition metal.Only one of the new large band gap materials B$_4$O$_6$ does not contain fluoride.

In Fig. \ref{fig:BS} we show the PBE and HSE06 band structures for four non-metallic compounds selected from the set of new monolayers: AgHgISe, MoF$_3$, AlF$_3$, and CrFO$_2$. All four materials have a direct band gap.

With a band gap of 10.8 eV (HSE06), AlF$_3$ has the largest band gap among all the new materials. Such large band gap 2D insulators are relevant for several reasons, including as gate dielectrics in field effect transistors employing 2D semiconductors as the channel material\cite{knobloch2021performance,xu2023scalable}. 
As an example of a simple, previously unknown monolayer we show the band structure of MoF$_3$. This material lies on the convex hull ($\Delta H_{\mathrm{hull}}=0$) and is predicted to have a direct band gap of 1.0 eV (HSE06), which is close to a widely used telecommunication wavelength band (O-band)\cite{Cao_2019}, making it interesting for photonics and opto-electronic applications in the visible frequency range. 
As an example of a more complex material we show the band structure of AgHgISe. This material has an energy slightly above the convex hull of ($\Delta H_{\mathrm{hull}}=0.02$ eV/atom) and a direct band gap of 1.88 eV (HSE06). 
Finally, as an example of a new magnetic material, we show the band structure of CrFO$_2$. This material also lies on the convex hull and has a direct band gap of 3.65 eV (HSE06).

\begin{figure*}
    \centering
    \includegraphics[]{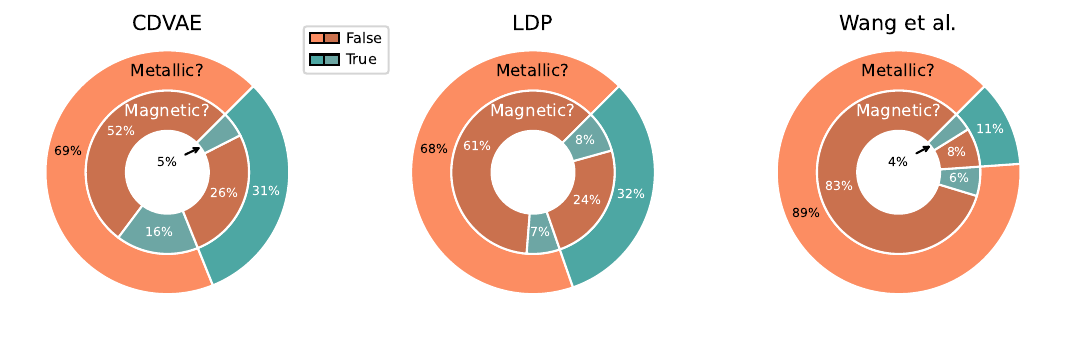}
    \caption{Distribution of the new materials according to their metallic/non-metallic and magnetic/non-magnetic nature.}
    \label{fig:metallic_magnetic}
\end{figure*}

\subsection{Magnetic materials}
All the DFT structure relaxations are performed with spin polarisation starting from an initial ferromagnetic spin configuration. If the final structure has absolute magnetic moments on all atoms below $0.1\mu_B$, the materials is considered non-magnetic and all subsequent property evaluations are performed on-top of a spin-paired ground state calculation. For magnetic materials, a nearest neighbor exchange coupling, $J$, is derived from the total energy of one specific anti-ferromagnetic configuration. If $J>0$, the material is classified as 'ferromagnetic'. If $J<0$, the material is classified as 'magnetic'. Note that in the case of $J<0$, we do not classify the material as 'anti-ferromagnetic', because (1) we have only considered one of many possible anti-ferromagnetic configurations and (2) the true magnetic ground state could be more complex, e.g. a spin spiral.

Fig. \ref{fig:metallic_magnetic} shows three pie charts depicting the fraction of new materials that are magnetic/non-magnetic and metallic/insulating. Magnetic materials with semiconducting properties are of particular interest for spintronics applications, e.g. spin transistors\cite{datta1990electronic}. Interestingly, the CDVAE-generated structures contain a significantly larger fraction of magnetic, non-metallic materials (16\%) than both the LDP (7\%) and Wang \emph{et al.} (6\%) structures.  

\begin{figure}
    \centering
    \includegraphics[]{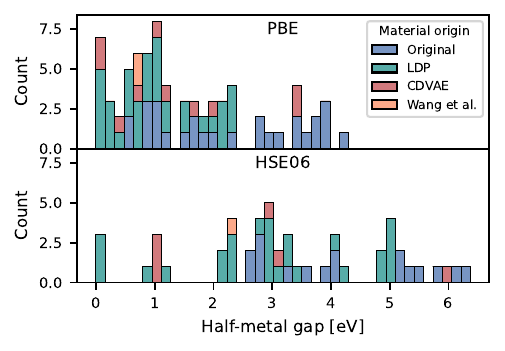}
    \caption{Histogram showing the distribution of half-metal gaps for materials which are magnetic, metallic, and have a gap in one spin channel. The half-metal gaps are calculated using PBE and HSE06, both without spin orbit coupling.}
    \label{fig:halfmetal}
\end{figure}

Another interesting class of materials within spintronics are half-metals, which are metallic in one of the spin channels while the other has a band gap\cite{ashton2017two}.  Calculating the half-metal gap was not previously part of the C2DB workflow but we have recently added it and calculated the half-metal gap for all monolayers in C2DB, which are magnetic, metallic, and dynamically stable. In total 254 monolayers satisfy these conditions and of these 74 have a non-zero gap in one of the spin-channels when using the PBE xc-functional. Among these 74 monolayers demonstrating a PBE-based half-metallic gap, further analysis was conducted employing the more accurate HSE06 xc-functional. In total 51 of these monolayers remains half-metallic at the HSE06 level. We note that spin orbit coupling (SOC) may mix the two spin channels making the very concept of a half-metal somewhat blurred. For this reason the calculation of the half-metal gap does not include SOC. The distribution of the half-metal gaps for both PBE and HSE06 is show in Fig. \ref{fig:halfmetal}. The half-metallic band gap of three of the monolayers, namely FeCl$_2$, FeBr$_2$, and FeI$_2$, was previously investigated at the HSE06 level in Ref. \cite{ashton2017two}. Our findings align well with theirs; they reported the (HSE06) half-metallic band gap to be 6.4 eV, 5.5 eV, and 4.0 eV, respectively, while our calculations yield values of 6.1 eV, 5.2 eV, and 4.0 eV for the same materials.

\begin{figure}
    \centering
    \includegraphics[width=9cm]{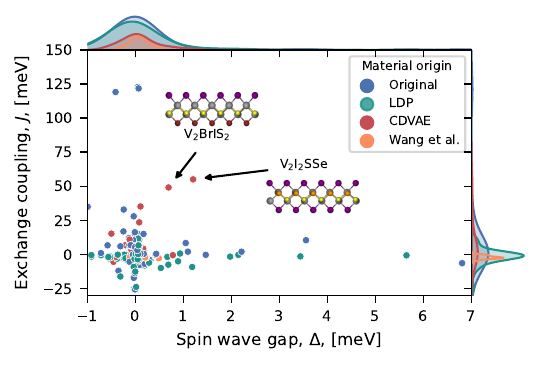}
    \caption{Nearest neighbor exchange coupling, $J$, versus the spin wave gap, $\Delta$, for the magnetic and non-metallic materials. Only materials with $\Delta >-1$ meV are shown, as a positive spin wave gap is a requirement for magnetic order in 2D. }
    \label{fig:delta_J}
\end{figure}

In addition to the total magnetic moment and the nearest neighbor exchange coupling, the magnetic anisotropy is also calculated by the workflow for materials with one or two magnetic atoms in the unit cell. It represents the total energy difference between spins aligned in the in-plane ($x$ and $y$) and out-of-plane ($z$) directions, respectively. Thus the sign of the magnetic anisotropy determines whether the magnetic material has an easy axis or easy plane. 

\begin{figure*}
    \centering
    \includegraphics[width=18.3cm]{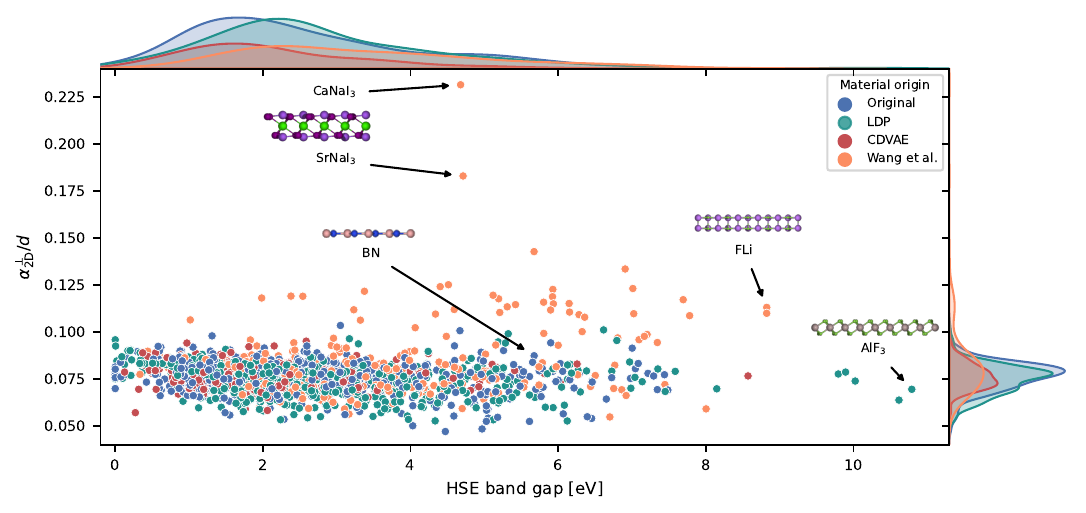}
    \caption{Static out-of-plane polarisability normalized by the estimated van der Waals thickness of the monolayer, $\alpha_{\mathrm{2D}}^\perp/d$, versus the HSE06 band gap (metals not included). Results are shown for the new materials from the different datasets and the original set of stable materials in the C2DB. The absence of any correlation between $\alpha_{\mathrm{2D}}$ and the size of the band gap is ascribed to two effects, namely, the contribution from infrared active phonons and the influence of local field effects, which is particularly strong for out-of-plane polarisation.}
    \label{fig:pol}
\end{figure*}

The magnetic properties of a material can be described using a Heisenberg model (assuming a single magnetic site per unit cell and nearest neighbor interactions)
\begin{equation}
    H = - \frac{J}{2} \sum_{i \neq j} \mathbf{S}_i \cdot \mathbf{S}_j - A \sum_i \left( S_i^z \right)^2 - \frac{B}{2} \sum_{i \neq j} S_i^z S_j^z.
\end{equation}
Here $J$ is the nearest neighbor exchange coupling, $A$ is the single-ion anisotropy (out-of-plane) and $B$ is the anisotropic exchange (out-of-plane). The sums over $i$ and $j$ is restricted to nearest neighbors. The parameters of the Heisenberg model can be obtained from DFT total energies as described in Ref.\cite{torelli2019high}.
Using spin wave theory, the spin wave gap, $\Delta$, i.e. the smallest energy required for a magnetic excitation, is given by\cite{torelli2019high}

\begin{equation}
    \Delta = A \left( 2 S - 1 \right) + B S N_{nn}
\end{equation}

where $N_{nn}$ is the number of nearest neighbors in the given crystal and $S$ is the spin. $\Delta > 0$ is required for for out-of-plane magnetism in 2D materials. By combining the exchange coupling and spin wave gap with simple structural parameters, e.g. number of nearest neighbors and lattice type, it is possible to estimate the critical temperature as described in Ref. \cite{torelli2019high}. In general for a high critical temperature, both the exchange coupling and the spin wave gap should be as large as possible.

In Fig. \ref{fig:delta_J} the exchange coupling, $J$, is plotted against the spin wave gap, $\Delta$, for the magnetic and non-metallic materials. The two CDVAE generated materials V$_2$I$_2$SSe and V$_2$BrIS$_2$ exhibit notably high values of $J$ and $\Delta$. These two share the same crystal structure and belong to layer group 11.

\subsection{Dielectric screening}
The frequency-dependent 2D optical polarisability, $\boldsymbol{\alpha}_{\mathrm{2D}}^{\mathrm{opt}}(\omega)$, in the long wave lenght limit ($\mathbf{q}\to \mathbf{0}$), is calculated within the random phase approximation (RPA). For metals, a Drude term, 
\begin{equation}
\alpha_{\mathrm{2D}}^{\mathrm{opt,intra}}(\omega) =-\frac{\omega_p^2}{2\pi\omega^2} 
\end{equation}
is added to the interband polarisability to account for intraband transitions. In this expression the plasma frequency is obtained as an integral of $\langle n\mathbf{k}|\hat{\mathbf{q}}\cdot \nabla |n\mathbf k\rangle$ over the Fermi surface, where $\hat{\mathbf{q}}$ is a unit vector in the direction of the electric field\cite{haastrup2018computational}

In materials with a finite band gap, the atomic lattice can also contribute to the polarisability at frequencies below or comparable to the maximum phonon frequency.
To include the contribution to the polarisability from infrared (IR) active phonons, we first determine the atomic Born charges describing the change in the macroscopic polarisation, $\mathbf{P}$, due to displacement of the atom, 
\begin{equation}
Z^a_{ij} = \partial P_i/\partial x^a_j|_{E=0}. 
\end{equation}
The IR polarisability can then be determined by combining the Born charges with the eigenmodes and frequencies of the optical phonons at the $\Gamma$-point. The total polarisability is then obtain as the sum
\begin{equation}
\boldsymbol{\alpha}_{\mathrm{2D}}^{\mathrm{tot}}(\omega)= \boldsymbol{\alpha}_{\mathrm{2D}}^{\mathrm{opt}}(\omega)+\boldsymbol{\alpha}_{\mathrm{2D}}^{\mathrm{IR}}(\omega)
\end{equation}
All quantities in the above equation are tensors. 

It has been proposed that the dielectric constant of a layered van der Waals (vdW) bulk crystal can be obtained from the polarisability of its constituent monolayers via the relations\cite{tian2019electronic}
\begin{eqnarray}\label{eq:eps_in}
    \epsilon_{\mathrm{bulk}}^{||} &=& 1 + 4\pi \frac{\alpha_{\mathrm{2D}}^{||}}{d} \\\label{eq:eps_out}
    \epsilon_{\mathrm{bulk}}^\perp &=& (1 - 4\pi \frac{\alpha_{\mathrm{2D}}^\perp}{d})^{-1}
\end{eqnarray}
where $d$ is taken as an effective thickness of the 2D material. We estimate $d$ as the distance between the two outermost atoms of the monolayer plus the vdW radii\cite{alvares2013a} of the outermost atoms. We have checked that this approximation gives a good qualitative agreement with the DFT calculated distance between monolayers in vdW bilayers, albeit slightly underestimating the thickness\cite{pakdel2023emergent}. 

As previously mentioned, 2D vdW materials with good dielectric properties are being actively sought for due to the potential application in 2D electronics, e.g. as gate insulators\cite{knobloch2021performance,osanloo2022transition,xu2023scalable}. Good field effect transistor gate dielectrics are characterised by a large electronic band gap (to limit leakage currents) and a large out-of-plane static dielectric constant (to minimise gate thickness and threshold voltage). To screen the new 2D materials for gate dielectric candidates we plot the fraction $\alpha_{\mathrm{2D}}^\perp / d$ against the HSE06 band gap, see Fig. \ref{fig:pol}. While our results show that Eq. (\ref{eq:eps_in}) is generally quite accurate, we have found that Eq. (\ref{eq:eps_out}) can lead to a diverging or even negative out-of-plane dielectric constant - even when the layer thickness ($d$) is derived from more accurate DFT calculations or experimental interlayer distances. For this reason we show $\alpha_{\mathrm{2D}}^\perp / d$, which directly expresses the ability of the individual 2D layer to screen an electric field.  We note in passing that the electronic contribution to the total static polarisability is expected to scale as $1/E_{\mathrm{gap}}$. While such a trend is indeed observed for the in-plane component ($\alpha_{\mathrm{2D}}^{||}$) it is almost absent for the out-of-plane component ($\alpha_{\mathrm{2D}}^\perp$). These observations agree with previous findings\cite{tian2019electronic}. We ascribe this different behavior to the larger influence of local field effects on the out-of-plane polarisability.  

A few of the new materials with particularly large values of the key quantity $E_{\mathrm{gap}} \alpha_{\mathrm{2D}}^\perp/d$ are highlighted and their atomic structure shown. In particular, CaNaI$_3$ and SrNaI$_3$, which shares structure prototype, have exceptionally high $\alpha_{\mathrm{2D}}^\perp / d$ and a reasonably large band gap of 4.4 eV. This 
is due to a very large phonon contribution to the polarisability, e.g. in the case of CaNaI$_3$, $\alpha_{\mathrm{2D}}^{\perp,\mathrm{IR}}=1.20$ \AA  while $\alpha_{\mathrm{2D}}^{\perp,\mathrm{opt}}=0.39$ \AA.
Hexagonal BN, which is commonly used in experimental studies, is also highlighted as is AlF$_3$ whose band structure is shown in Fig. \ref{fig:BS}. In general, the materials from Wang \emph{et al.} contain several candidates with large phonon contributions to $\alpha_{\mathrm{2D}}^\perp$.

\subsection{Piezoelectric tensor}

\begin{figure}
    \centering
    \includegraphics[width=9cm]{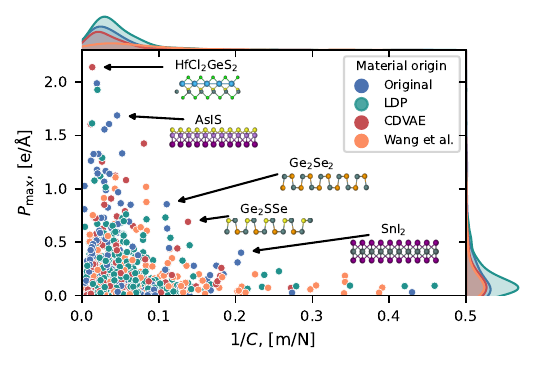}
    \caption{Maximum polarisation, $P_{\mathrm{max}}$, versus the inverse stiffness, $1/C$. The stiffness is calculated in the direction of the maximum polarisation.}
    \label{fig:piezo}
\end{figure}

The piezoelectric tensor has been calculated for 858 of the new materials that are dynamically stable, non-centrosymmetric, and have a finite band gap.

The piezoelectric tensor, $c$, of an insulating crystal is a rank-3 tensor relating the macroscopic polarization, $P$, to an applied strain. It is non-zero only for crystals lacking an inversion center. In Voigt notation, $c$ is expressed as a $3\times N$
matrix relating the $(x,y,z)$ components of the macroscopic polarizability to the $N$
independent components of the strain tensor. The piezoelectric tensor is evaluated as a finite difference of the polarization under three independent strains of the unit cell with the atom positions fully relaxed. The polarisation in the periodic directions is calculated as an integral over Berry phases. The polarization in the non-periodic direction is obtained by direct evaluation of the first moment of the electron density.

For 2D materials, the strain tensor has three independent components comprising two linear ($xx$ and $yy$) and one shear ($xy$) component. Thus strain can be represented as a 3-vector, and the Piezoelectric tensor as a $3\times 3$ matrix,
\begin{equation}
    c_{ij} = \frac{\partial P_i}{\partial \epsilon_j}
\end{equation}
where $i=x,y,z$ and $j=xx,yy,xy$. As a validation of the computational methodology we mention that the calculated piezoelectric coupling of freestanding MoS$_2$ is 0.35 nC/m in good agreement with the experimental value of 0.3 nC/m. See Ref. \cite{gjerding2021recent} for further details on the computational method. 

Piezoelectric 2D materials could find applications in nanoscale electro-mechanical devices, mechanical-electrical energy conversion, and sensing\cite{zhu2015observation,wu2014piezoelectricity}. 
For a high mechanical-electrical energy conversion, the strain-induced polarisation should be as large as possible, and the elastic energy as small as possible. The strain direction of maximum polarisation is the eigenvector ($\mathbf{e}_{\mathrm{max}}$) corresponding to the largest eigenvalue of the matrix $c^\dagger c$.
The stiffness of the material in this direction is then $\mathbf{e}_{\mathrm{max}}^\dagger C \mathbf{e}_{\mathrm{max}}$.
In Fig. \ref{fig:piezo} we plot the maximum polarisation against the inverse stiffness along the direction of the maximum polarisation for the different materials groups. Some of the materials that look particularly promising mechanical-electrical energy conversion, are highlighted.

\begin{figure*}
    \centering
    \includegraphics[]{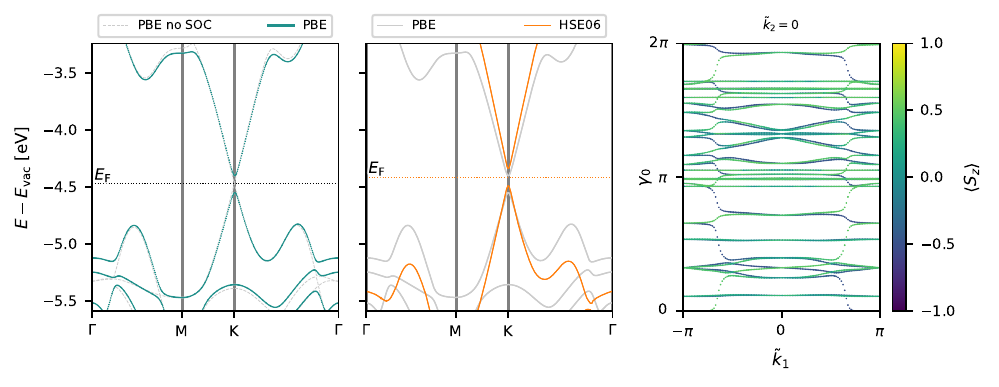}
    \caption{Electronic band structure calculated with the PBE (left) and HSE06 (middle) xc-functionals, and the Berry phase spectrum (right) of the CDVAE-generated monolayer Ta$_2$Te$_2$S. The material is a quantum spin Hall insulator ($Z_2$ index $\nu=1$) with a HSE06 band gap of 0.15 eV at the K-point.  }
    \label{fig:berryQSH}
\end{figure*}

\begin{figure*}
    \centering
    \includegraphics[]{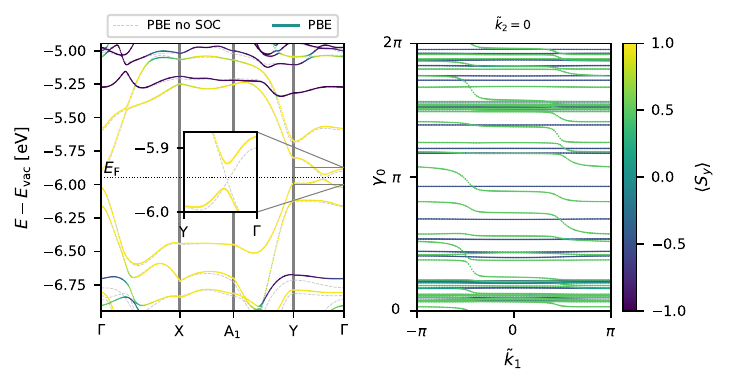}
    \caption{Electronic band structure and Berry phase spectrum of the CDVAE-generated monolayer Mn$_2$Br$_2$O$_3$ obtained using the PBE xc-functional. The material is ferromagnetic and is classified as an anomalous Hall insulator (Chern number $C=1$) with a PBE band gap of 0.04 eV. }
    \label{fig:berryAH}
\end{figure*}

\subsection{Topological invariants}
The Berry phase spectrum is calculated for 340 of the new materials with a band gap in the range $0<E^\mathrm{PBE}_{\mathrm{gap}}<0.7$ eV and with less than 13 atoms in the unit cell. The Berry phase spectrum gives the Berry phase, $\gamma_n(\Tilde{k_1})$, of an occupied band, $n$, along the path $\Tilde{k_2}=0\rightarrow \Tilde{k_2}=2\pi$, where a specific $k$ point in reciprocal space is given by $(\Tilde{k_1}/{2\pi}) \mathbf{b}_1+(\Tilde{k_2}/{2\pi})\mathbf{b}_2$. The topological indices, in particular the Chern number ($C$), the mirror Chern number ($C_M$), and the $Z_2$ invariant ($\nu$), can be determined by inspection of the Berry phase spectrum. We refer to Refs. \cite{gjerding2021atomic} and \cite{olsen2019discovering} for more details on the methodology.

An example of a quantum spin Hall insulator is the CDVAE-generated Ta$_2$Te$_2$S. The Berry phase spectrum of this materials is shown in Fig. \ref{fig:berryQSH} (right) while the electronic band structures calculated with PBE and HSE06 are shown in the left and middle panels, respectively.  From the band structure we note that Ta$_2$Te$_2$S, like graphene\cite{neto2009the} and silicene\cite{kharadi2020review}, hosts a Dirac cone at the K point. However, the band gap of Ta$_2$Te$_2$S (150 meV) is much larger than that of pristine graphene (no gap) and silicene (1.55 meV \cite{liu2011quantum}), making it more suitable for applications.

The only example among the new materials of an anomalous Hall insulator is the CDVAE-generated magnetic monolayer Mn$_2$Br$_2$O$_3$. Its Berry phase spectrum and PBE electronic band structure is shown in Fig. \ref{fig:berryAH}.

\section{Summary}
We have performed an extensive computational characterisation of 2759 two-dimensional (2D) crystals all of which are predicted to be dynamically and thermodynamically stable, but never have been explored before. Using state of the art \emph{ab initio} calculations we have determined a variety of basic properties of this previously unknown set materials and made them available in the Computational 2D Materials Database (C2DB). As a result, this work increases the number of stable materials contained in the C2DB by almost a factor of three.

We identify 633 monolayers with a band gap in the semiconductor energy range 0.5-2 eV, of which 156 are direct gaps (results obtained with the HSE06 xc-functional including spin-orbit coupling). We find 406 materials with a magnetic ground state of which 216 have a finite band gap. Among these several exhibit a large magnetic anisotropy, which is an essential requirement for magnetic order at finite temperatures in the 2D limit. In particular, we find a number of semiconducting materials (e.g. the chalcohalides V$_2$BrIS$_2$ and V$_2$I$_2$SSe) with simultaneously large exchange coupling and spin wave gaps making them candidates for 2D magnets with high Curie temperature.   

Our calculations provide a complete characterisation of the linear dielectric properties of the non-metallic monolayers. Specifically, both the electronic and phononic contributions to the polarisability are calculated as function of frequency for in-plane and out-of-plane polarisation directions. Focusing on the static out-of-plane polarisability, which is of greatest relevance for dielectric applications in 2D electronics, we identify a number of promising materials combining a high polarisability with a large band gap, e.g. CaNaI$_3$ and SrNaI$_3$. Not surprisingly, these materials are characterised by a relatively large contribution to the polarisability from out-of-plane optical phonons.

The piezoelectric tensor is calculated for more than 800 non-centrosymmetric and insulating monolayers. Based on a combined analysis of the piezoelectric tensor and the stiffness tensor, we identified a number of promising materials for mechanical-electrical energy conversion. 

Finally, we have calculated the Berry phase spectrum of 340 materials with small band gaps and used to identify materials with topologically non-trivial band structures. Here we highlighted Ta$_2$Te$_2$S, which we predict to be a quantum spin Hall insulator with a gapped Dirac cone, and Mn$_2$Br$_2$O$_3$, which is classified as a ferromagnetic anomalous Hall insulator.

\section{Method} \label{sec: Method}
The employed C2DB computational workflow was constructed within the Atomic Simulation Recipes (ASR) Python framework\cite{gjerding2021atomic} and executed using the MyQueue\cite{mortensen2020myqueue} task scheduler. The workflow performs calculations using the GPAW\cite{enkovaara2010electronic} electronic structure code and the Atomic Simulation Environment (ASE) Python library\cite{larsen2017atomic}.

All GPAW DFT calculations were performed using an 800 eV plane wave cut-off and a k-point grids with density between 4 and 12 \AA depending on the property calculated. The PBE exchange-correlation functional\cite{perdew1996generalized} was used in all calculations, except for the HSE06\cite{heyd2003hybrid} band structure calculations. Spin-orbit coupling was included in calculations of single-particle band energies and the magnetic anisotropy. An out-of-plane vacuum region of at least 12 Å is included in the calculations to eliminate any spurious interactions between the periodic images of the 2D layers. All the property calculations has previously been shown to be well converged for the given computational setting used \cite{haastrup2018computational, gjerding2021recent}. The precise computational settings used for each type of property calculation can be found in Ref. \cite{gjerding2021recent}.

\section{Data availability}
The data that support the findings of this study are openly available. All the crystal structures and their properties are available as a part of C2DB (\url{https://cmr.fysik.dtu.dk/c2db/c2db.html})

\section{Acknowledgements}
The authors acknowledge funding from the European Research Council (ERC) under the European Union’s Horizon 2020 research and innovation program Grant No. 773122 (LIMA) and Grant agreement No. 951786 (NOMAD CoE). K. S. T. acknowledges support from the Novo Nordisk Foundation Challenge Programme 2021: Smart nanomaterials for applications in life-science, BIOMAG Grant No. NNF21OC0066526.
K. S. T. is a Villum Investigator supported by VILLUM FONDEN (grant no. 37789).

\section{Competing interests}
The authors declare no competing interests.

\section{Author contributions}
P.L. and K.S.T. developed the initial concept. P.L. ran the DFT simulations and performed the data analysis. K.S.T. supervised the project and aided with the interpretation of the results. P.L. and K.S.T wrote and discussed the paper together.

\bibliography{refs}

\begin{thebibliography}{46}%
\makeatletter
\providecommand \@ifxundefined [1]{%
 \@ifx{#1\undefined}
}%
\providecommand \@ifnum [1]{%
 \ifnum #1\expandafter \@firstoftwo
 \else \expandafter \@secondoftwo
 \fi
}%
\providecommand \@ifx [1]{%
 \ifx #1\expandafter \@firstoftwo
 \else \expandafter \@secondoftwo
 \fi
}%
\providecommand \natexlab [1]{#1}%
\providecommand \enquote  [1]{``#1''}%
\providecommand \bibnamefont  [1]{#1}%
\providecommand \bibfnamefont [1]{#1}%
\providecommand \citenamefont [1]{#1}%
\providecommand \href@noop [0]{\@secondoftwo}%
\providecommand \href [0]{\begingroup \@sanitize@url \@href}%
\providecommand \@href[1]{\@@startlink{#1}\@@href}%
\providecommand \@@href[1]{\endgroup#1\@@endlink}%
\providecommand \@sanitize@url [0]{\catcode `\\12\catcode `\$12\catcode `\&12\catcode `\#12\catcode `\^12\catcode `\_12\catcode `\%12\relax}%
\providecommand \@@startlink[1]{}%
\providecommand \@@endlink[0]{}%
\providecommand \url  [0]{\begingroup\@sanitize@url \@url }%
\providecommand \@url [1]{\endgroup\@href {#1}{\urlprefix }}%
\providecommand \urlprefix  [0]{URL }%
\providecommand \Eprint [0]{\href }%
\providecommand \doibase [0]{http://dx.doi.org/}%
\providecommand \selectlanguage [0]{\@gobble}%
\providecommand \bibinfo  [0]{\@secondoftwo}%
\providecommand \bibfield  [0]{\@secondoftwo}%
\providecommand \translation [1]{[#1]}%
\providecommand \BibitemOpen [0]{}%
\providecommand \bibitemStop [0]{}%
\providecommand \bibitemNoStop [0]{.\EOS\space}%
\providecommand \EOS [0]{\spacefactor3000\relax}%
\providecommand \BibitemShut  [1]{\csname bibitem#1\endcsname}%
\let\auto@bib@innerbib\@empty
\bibitem [{\citenamefont {Marzari}\ \emph {et~al.}(2021)\citenamefont {Marzari}, \citenamefont {Ferretti},\ and\ \citenamefont {Wolverton}}]{marzari2021electronic}%
  \BibitemOpen
  \bibfield  {author} {\bibinfo {author} {\bibfnamefont {N.}~\bibnamefont {Marzari}}, \bibinfo {author} {\bibfnamefont {A.}~\bibnamefont {Ferretti}}, \ and\ \bibinfo {author} {\bibfnamefont {C.}~\bibnamefont {Wolverton}},\ }\href@noop {} {\bibfield  {journal} {\bibinfo  {journal} {Nature Materials}\ }\textbf {\bibinfo {volume} {20}},\ \bibinfo {pages} {736} (\bibinfo {year} {2021})}\BibitemShut {NoStop}%
\bibitem [{\citenamefont {Thygesen}\ and\ \citenamefont {Jacobsen}(2016)}]{thygesen2016making}%
  \BibitemOpen
  \bibfield  {author} {\bibinfo {author} {\bibfnamefont {K.~S.}\ \bibnamefont {Thygesen}}\ and\ \bibinfo {author} {\bibfnamefont {K.~W.}\ \bibnamefont {Jacobsen}},\ }\href@noop {} {\bibfield  {journal} {\bibinfo  {journal} {Science}\ }\textbf {\bibinfo {volume} {354}},\ \bibinfo {pages} {180} (\bibinfo {year} {2016})}\BibitemShut {NoStop}%
\bibitem [{\citenamefont {Saal}\ \emph {et~al.}(2013)\citenamefont {Saal}, \citenamefont {Kirklin}, \citenamefont {Aykol}, \citenamefont {Meredig},\ and\ \citenamefont {Wolverton}}]{saal2013materials}%
  \BibitemOpen
  \bibfield  {author} {\bibinfo {author} {\bibfnamefont {J.~E.}\ \bibnamefont {Saal}}, \bibinfo {author} {\bibfnamefont {S.}~\bibnamefont {Kirklin}}, \bibinfo {author} {\bibfnamefont {M.}~\bibnamefont {Aykol}}, \bibinfo {author} {\bibfnamefont {B.}~\bibnamefont {Meredig}}, \ and\ \bibinfo {author} {\bibfnamefont {C.}~\bibnamefont {Wolverton}},\ }\href@noop {} {\bibfield  {journal} {\bibinfo  {journal} {{JOM}}\ }\textbf {\bibinfo {volume} {65}},\ \bibinfo {pages} {1501} (\bibinfo {year} {2013})}\BibitemShut {NoStop}%
\bibitem [{\citenamefont {Curtarolo}\ \emph {et~al.}(2012)\citenamefont {Curtarolo}, \citenamefont {Setyawan}, \citenamefont {Hart}, \citenamefont {Jahnatek}, \citenamefont {Chepulskii}, \citenamefont {Taylor}, \citenamefont {Wang}, \citenamefont {Xue}, \citenamefont {Yang}, \citenamefont {Levy} \emph {et~al.}}]{curtarolo2012aflow}%
  \BibitemOpen
  \bibfield  {author} {\bibinfo {author} {\bibfnamefont {S.}~\bibnamefont {Curtarolo}}, \bibinfo {author} {\bibfnamefont {W.}~\bibnamefont {Setyawan}}, \bibinfo {author} {\bibfnamefont {G.~L.}\ \bibnamefont {Hart}}, \bibinfo {author} {\bibfnamefont {M.}~\bibnamefont {Jahnatek}}, \bibinfo {author} {\bibfnamefont {R.~V.}\ \bibnamefont {Chepulskii}}, \bibinfo {author} {\bibfnamefont {R.~H.}\ \bibnamefont {Taylor}}, \bibinfo {author} {\bibfnamefont {S.}~\bibnamefont {Wang}}, \bibinfo {author} {\bibfnamefont {J.}~\bibnamefont {Xue}}, \bibinfo {author} {\bibfnamefont {K.}~\bibnamefont {Yang}}, \bibinfo {author} {\bibfnamefont {O.}~\bibnamefont {Levy}},  \emph {et~al.},\ }\href@noop {} {\bibfield  {journal} {\bibinfo  {journal} {Computational Materials Science}\ }\textbf {\bibinfo {volume} {58}},\ \bibinfo {pages} {218} (\bibinfo {year} {2012})}\BibitemShut {NoStop}%
\bibitem [{\citenamefont {Montavon}\ \emph {et~al.}(2013)\citenamefont {Montavon}, \citenamefont {Rupp}, \citenamefont {Gobre}, \citenamefont {Vazquez-Mayagoitia}, \citenamefont {Hansen}, \citenamefont {Tkatchenko}, \citenamefont {M{\"u}ller},\ and\ \citenamefont {Von~Lilienfeld}}]{montavon2013machine}%
  \BibitemOpen
  \bibfield  {author} {\bibinfo {author} {\bibfnamefont {G.}~\bibnamefont {Montavon}}, \bibinfo {author} {\bibfnamefont {M.}~\bibnamefont {Rupp}}, \bibinfo {author} {\bibfnamefont {V.}~\bibnamefont {Gobre}}, \bibinfo {author} {\bibfnamefont {A.}~\bibnamefont {Vazquez-Mayagoitia}}, \bibinfo {author} {\bibfnamefont {K.}~\bibnamefont {Hansen}}, \bibinfo {author} {\bibfnamefont {A.}~\bibnamefont {Tkatchenko}}, \bibinfo {author} {\bibfnamefont {K.-R.}\ \bibnamefont {M{\"u}ller}}, \ and\ \bibinfo {author} {\bibfnamefont {O.~A.}\ \bibnamefont {Von~Lilienfeld}},\ }\href@noop {} {\bibfield  {journal} {\bibinfo  {journal} {New Journal of Physics}\ }\textbf {\bibinfo {volume} {15}},\ \bibinfo {pages} {095003} (\bibinfo {year} {2013})}\BibitemShut {NoStop}%
\bibitem [{\citenamefont {Ward}\ \emph {et~al.}(2016)\citenamefont {Ward}, \citenamefont {Agrawal}, \citenamefont {Choudhary},\ and\ \citenamefont {Wolverton}}]{ward2016general}%
  \BibitemOpen
  \bibfield  {author} {\bibinfo {author} {\bibfnamefont {L.}~\bibnamefont {Ward}}, \bibinfo {author} {\bibfnamefont {A.}~\bibnamefont {Agrawal}}, \bibinfo {author} {\bibfnamefont {A.}~\bibnamefont {Choudhary}}, \ and\ \bibinfo {author} {\bibfnamefont {C.}~\bibnamefont {Wolverton}},\ }\href@noop {} {\bibfield  {journal} {\bibinfo  {journal} {npj Computational Materials}\ }\textbf {\bibinfo {volume} {2}},\ \bibinfo {pages} {1} (\bibinfo {year} {2016})}\BibitemShut {NoStop}%
\bibitem [{\citenamefont {Manti}\ \emph {et~al.}(2023)\citenamefont {Manti}, \citenamefont {Svendsen}, \citenamefont {Kn{\o}sgaard}, \citenamefont {Lyngby},\ and\ \citenamefont {Thygesen}}]{manti2023exploring}%
  \BibitemOpen
  \bibfield  {author} {\bibinfo {author} {\bibfnamefont {S.}~\bibnamefont {Manti}}, \bibinfo {author} {\bibfnamefont {M.~K.}\ \bibnamefont {Svendsen}}, \bibinfo {author} {\bibfnamefont {N.~R.}\ \bibnamefont {Kn{\o}sgaard}}, \bibinfo {author} {\bibfnamefont {P.~M.}\ \bibnamefont {Lyngby}}, \ and\ \bibinfo {author} {\bibfnamefont {K.~S.}\ \bibnamefont {Thygesen}},\ }\href@noop {} {\bibfield  {journal} {\bibinfo  {journal} {npj Computational Materials}\ }\textbf {\bibinfo {volume} {9}},\ \bibinfo {pages} {33} (\bibinfo {year} {2023})}\BibitemShut {NoStop}%
\bibitem [{\citenamefont {Wei}\ \emph {et~al.}(2019)\citenamefont {Wei}, \citenamefont {Chu}, \citenamefont {Sun}, \citenamefont {Xu}, \citenamefont {Deng}, \citenamefont {Chen}, \citenamefont {Wei},\ and\ \citenamefont {Lei}}]{wei2019machine}%
  \BibitemOpen
  \bibfield  {author} {\bibinfo {author} {\bibfnamefont {J.}~\bibnamefont {Wei}}, \bibinfo {author} {\bibfnamefont {X.}~\bibnamefont {Chu}}, \bibinfo {author} {\bibfnamefont {X.-Y.}\ \bibnamefont {Sun}}, \bibinfo {author} {\bibfnamefont {K.}~\bibnamefont {Xu}}, \bibinfo {author} {\bibfnamefont {H.-X.}\ \bibnamefont {Deng}}, \bibinfo {author} {\bibfnamefont {J.}~\bibnamefont {Chen}}, \bibinfo {author} {\bibfnamefont {Z.}~\bibnamefont {Wei}}, \ and\ \bibinfo {author} {\bibfnamefont {M.}~\bibnamefont {Lei}},\ }\href@noop {} {\bibfield  {journal} {\bibinfo  {journal} {InfoMat}\ }\textbf {\bibinfo {volume} {1}},\ \bibinfo {pages} {338} (\bibinfo {year} {2019})}\BibitemShut {NoStop}%
\bibitem [{\citenamefont {Mueller}\ \emph {et~al.}(2016)\citenamefont {Mueller}, \citenamefont {Kusne},\ and\ \citenamefont {Ramprasad}}]{mueller2016machine}%
  \BibitemOpen
  \bibfield  {author} {\bibinfo {author} {\bibfnamefont {T.}~\bibnamefont {Mueller}}, \bibinfo {author} {\bibfnamefont {A.~G.}\ \bibnamefont {Kusne}}, \ and\ \bibinfo {author} {\bibfnamefont {R.}~\bibnamefont {Ramprasad}},\ }\href@noop {} {\bibfield  {journal} {\bibinfo  {journal} {Reviews in Computational Chemistry}\ }\textbf {\bibinfo {volume} {29}},\ \bibinfo {pages} {186} (\bibinfo {year} {2016})}\BibitemShut {NoStop}%
\bibitem [{\citenamefont {Himanen}\ \emph {et~al.}(2019)\citenamefont {Himanen}, \citenamefont {Geurts}, \citenamefont {Foster},\ and\ \citenamefont {Rinke}}]{himanen2019data}%
  \BibitemOpen
  \bibfield  {author} {\bibinfo {author} {\bibfnamefont {L.}~\bibnamefont {Himanen}}, \bibinfo {author} {\bibfnamefont {A.}~\bibnamefont {Geurts}}, \bibinfo {author} {\bibfnamefont {A.~S.}\ \bibnamefont {Foster}}, \ and\ \bibinfo {author} {\bibfnamefont {P.}~\bibnamefont {Rinke}},\ }\href@noop {} {\bibfield  {journal} {\bibinfo  {journal} {Advanced Science}\ }\textbf {\bibinfo {volume} {6}},\ \bibinfo {pages} {1900808} (\bibinfo {year} {2019})}\BibitemShut {NoStop}%
\bibitem [{\citenamefont {Schmidt}\ \emph {et~al.}(2019)\citenamefont {Schmidt}, \citenamefont {Marques}, \citenamefont {Botti},\ and\ \citenamefont {Marques}}]{schmidt2019recent}%
  \BibitemOpen
  \bibfield  {author} {\bibinfo {author} {\bibfnamefont {J.}~\bibnamefont {Schmidt}}, \bibinfo {author} {\bibfnamefont {M.~R.}\ \bibnamefont {Marques}}, \bibinfo {author} {\bibfnamefont {S.}~\bibnamefont {Botti}}, \ and\ \bibinfo {author} {\bibfnamefont {M.~A.}\ \bibnamefont {Marques}},\ }\href@noop {} {\bibfield  {journal} {\bibinfo  {journal} {npj Computational Materials}\ }\textbf {\bibinfo {volume} {5}},\ \bibinfo {pages} {83} (\bibinfo {year} {2019})}\BibitemShut {NoStop}%
\bibitem [{\citenamefont {Kn{\o}sgaard}\ and\ \citenamefont {Thygesen}(2022)}]{knosgaard2022representing}%
  \BibitemOpen
  \bibfield  {author} {\bibinfo {author} {\bibfnamefont {N.~R.}\ \bibnamefont {Kn{\o}sgaard}}\ and\ \bibinfo {author} {\bibfnamefont {K.~S.}\ \bibnamefont {Thygesen}},\ }\href@noop {} {\bibfield  {journal} {\bibinfo  {journal} {Nature Communications}\ }\textbf {\bibinfo {volume} {13}},\ \bibinfo {pages} {468} (\bibinfo {year} {2022})}\BibitemShut {NoStop}%
\bibitem [{\citenamefont {Merchant}\ \emph {et~al.}(2023)\citenamefont {Merchant}, \citenamefont {Batzner}, \citenamefont {Schoenholz}, \citenamefont {Aykol}, \citenamefont {Cheon},\ and\ \citenamefont {Cubuk}}]{merchant2023novel}%
  \BibitemOpen
  \bibfield  {author} {\bibinfo {author} {\bibfnamefont {A.}~\bibnamefont {Merchant}}, \bibinfo {author} {\bibfnamefont {S.}~\bibnamefont {Batzner}}, \bibinfo {author} {\bibfnamefont {S.~S.}\ \bibnamefont {Schoenholz}}, \bibinfo {author} {\bibfnamefont {M.}~\bibnamefont {Aykol}}, \bibinfo {author} {\bibfnamefont {G.}~\bibnamefont {Cheon}}, \ and\ \bibinfo {author} {\bibfnamefont {E.~D.}\ \bibnamefont {Cubuk}},\ }\href {\doibase 10.1038/s41586-023-06735-9} {\bibfield  {journal} {\bibinfo  {journal} {Nature}\ }\textbf {\bibinfo {volume} {624}},\ \bibinfo {pages} {80} (\bibinfo {year} {2023})}\BibitemShut {NoStop}%
\bibitem [{\citenamefont {Schmidt}\ \emph {et~al.}(2022{\natexlab{a}})\citenamefont {Schmidt}, \citenamefont {Wang}, \citenamefont {Cerqueira}, \citenamefont {Botti},\ and\ \citenamefont {Marques}}]{schmidt2022a}%
  \BibitemOpen
  \bibfield  {author} {\bibinfo {author} {\bibfnamefont {J.}~\bibnamefont {Schmidt}}, \bibinfo {author} {\bibfnamefont {H.-C.}\ \bibnamefont {Wang}}, \bibinfo {author} {\bibfnamefont {T.~F.~T.}\ \bibnamefont {Cerqueira}}, \bibinfo {author} {\bibfnamefont {S.}~\bibnamefont {Botti}}, \ and\ \bibinfo {author} {\bibfnamefont {M.~A.~L.}\ \bibnamefont {Marques}},\ }\href {\doibase 10.1038/s41597-022-01177-w} {\bibfield  {journal} {\bibinfo  {journal} {Scientific Data}\ }\textbf {\bibinfo {volume} {9}},\ \bibinfo {pages} {64} (\bibinfo {year} {2022}{\natexlab{a}})}\BibitemShut {NoStop}%
\bibitem [{\citenamefont {Schmidt}\ \emph {et~al.}(2022{\natexlab{b}})\citenamefont {Schmidt}, \citenamefont {Hoffmann}, \citenamefont {Wang}, \citenamefont {Borlido}, \citenamefont {Carriço}, \citenamefont {Cerqueira}, \citenamefont {Botti},\ and\ \citenamefont {Marques}}]{schmidt2022largescale}%
  \BibitemOpen
  \bibfield  {author} {\bibinfo {author} {\bibfnamefont {J.}~\bibnamefont {Schmidt}}, \bibinfo {author} {\bibfnamefont {N.}~\bibnamefont {Hoffmann}}, \bibinfo {author} {\bibfnamefont {H.-C.}\ \bibnamefont {Wang}}, \bibinfo {author} {\bibfnamefont {P.}~\bibnamefont {Borlido}}, \bibinfo {author} {\bibfnamefont {P.~J. M.~A.}\ \bibnamefont {Carriço}}, \bibinfo {author} {\bibfnamefont {T.~F.~T.}\ \bibnamefont {Cerqueira}}, \bibinfo {author} {\bibfnamefont {S.}~\bibnamefont {Botti}}, \ and\ \bibinfo {author} {\bibfnamefont {M.~A.~L.}\ \bibnamefont {Marques}},\ }\href@noop {} {\enquote {\bibinfo {title} {Large-scale machine-learning-assisted exploration of the whole materials space},}\ } (\bibinfo {year} {2022}{\natexlab{b}}),\ \Eprint {http://arxiv.org/abs/2210.00579} {arXiv:2210.00579 [cond-mat.mtrl-sci]} \BibitemShut {NoStop}%
\bibitem [{\citenamefont {Ashton}\ \emph {et~al.}(2017{\natexlab{a}})\citenamefont {Ashton}, \citenamefont {Paul}, \citenamefont {Sinnott},\ and\ \citenamefont {Hennig}}]{ashton2017topology}%
  \BibitemOpen
  \bibfield  {author} {\bibinfo {author} {\bibfnamefont {M.}~\bibnamefont {Ashton}}, \bibinfo {author} {\bibfnamefont {J.}~\bibnamefont {Paul}}, \bibinfo {author} {\bibfnamefont {S.~B.}\ \bibnamefont {Sinnott}}, \ and\ \bibinfo {author} {\bibfnamefont {R.~G.}\ \bibnamefont {Hennig}},\ }\href@noop {} {\bibfield  {journal} {\bibinfo  {journal} {Physical Review Letters}\ }\textbf {\bibinfo {volume} {118}},\ \bibinfo {pages} {106101} (\bibinfo {year} {2017}{\natexlab{a}})}\BibitemShut {NoStop}%
\bibitem [{\citenamefont {Mounet}\ \emph {et~al.}(2018)\citenamefont {Mounet}, \citenamefont {Gibertini}, \citenamefont {Schwaller}, \citenamefont {Campi}, \citenamefont {Merkys}, \citenamefont {Marrazzo}, \citenamefont {Sohier}, \citenamefont {Castelli}, \citenamefont {Cepellotti}, \citenamefont {Pizzi},\ and\ \citenamefont {Marzari}}]{mounet2018two}%
  \BibitemOpen
  \bibfield  {author} {\bibinfo {author} {\bibfnamefont {N.}~\bibnamefont {Mounet}}, \bibinfo {author} {\bibfnamefont {M.}~\bibnamefont {Gibertini}}, \bibinfo {author} {\bibfnamefont {P.}~\bibnamefont {Schwaller}}, \bibinfo {author} {\bibfnamefont {D.}~\bibnamefont {Campi}}, \bibinfo {author} {\bibfnamefont {A.}~\bibnamefont {Merkys}}, \bibinfo {author} {\bibfnamefont {A.}~\bibnamefont {Marrazzo}}, \bibinfo {author} {\bibfnamefont {T.}~\bibnamefont {Sohier}}, \bibinfo {author} {\bibfnamefont {I.~E.}\ \bibnamefont {Castelli}}, \bibinfo {author} {\bibfnamefont {A.}~\bibnamefont {Cepellotti}}, \bibinfo {author} {\bibfnamefont {G.}~\bibnamefont {Pizzi}}, \ and\ \bibinfo {author} {\bibfnamefont {N.}~\bibnamefont {Marzari}},\ }\href@noop {} {\bibfield  {journal} {\bibinfo  {journal} {Nature Nanotechnology}\ }\textbf {\bibinfo {volume} {13}},\ \bibinfo {pages} {246} (\bibinfo {year} {2018})}\BibitemShut {NoStop}%
\bibitem [{\citenamefont {Haastrup}\ \emph {et~al.}(2018)\citenamefont {Haastrup}, \citenamefont {Strange}, \citenamefont {Pandey}, \citenamefont {Deilmann}, \citenamefont {Schmidt}, \citenamefont {Hinsche}, \citenamefont {Gjerding}, \citenamefont {Torelli}, \citenamefont {Larsen}, \citenamefont {Riis-Jensen} \emph {et~al.}}]{haastrup2018computational}%
  \BibitemOpen
  \bibfield  {author} {\bibinfo {author} {\bibfnamefont {S.}~\bibnamefont {Haastrup}}, \bibinfo {author} {\bibfnamefont {M.}~\bibnamefont {Strange}}, \bibinfo {author} {\bibfnamefont {M.}~\bibnamefont {Pandey}}, \bibinfo {author} {\bibfnamefont {T.}~\bibnamefont {Deilmann}}, \bibinfo {author} {\bibfnamefont {P.~S.}\ \bibnamefont {Schmidt}}, \bibinfo {author} {\bibfnamefont {N.~F.}\ \bibnamefont {Hinsche}}, \bibinfo {author} {\bibfnamefont {M.~N.}\ \bibnamefont {Gjerding}}, \bibinfo {author} {\bibfnamefont {D.}~\bibnamefont {Torelli}}, \bibinfo {author} {\bibfnamefont {P.~M.}\ \bibnamefont {Larsen}}, \bibinfo {author} {\bibfnamefont {A.~C.}\ \bibnamefont {Riis-Jensen}},  \emph {et~al.},\ }\href@noop {} {\bibfield  {journal} {\bibinfo  {journal} {2D Materials}\ }\textbf {\bibinfo {volume} {5}},\ \bibinfo {pages} {042002} (\bibinfo {year} {2018})}\BibitemShut {NoStop}%
\bibitem [{\citenamefont {Gjerding}\ \emph {et~al.}(2021{\natexlab{a}})\citenamefont {Gjerding}, \citenamefont {Skovhus}, \citenamefont {Rasmussen}, \citenamefont {Bertoldo}, \citenamefont {Larsen}, \citenamefont {Mortensen},\ and\ \citenamefont {Thygesen}}]{gjerding2021atomic}%
  \BibitemOpen
  \bibfield  {author} {\bibinfo {author} {\bibfnamefont {M.}~\bibnamefont {Gjerding}}, \bibinfo {author} {\bibfnamefont {T.}~\bibnamefont {Skovhus}}, \bibinfo {author} {\bibfnamefont {A.}~\bibnamefont {Rasmussen}}, \bibinfo {author} {\bibfnamefont {F.}~\bibnamefont {Bertoldo}}, \bibinfo {author} {\bibfnamefont {A.~H.}\ \bibnamefont {Larsen}}, \bibinfo {author} {\bibfnamefont {J.~J.}\ \bibnamefont {Mortensen}}, \ and\ \bibinfo {author} {\bibfnamefont {K.~S.}\ \bibnamefont {Thygesen}},\ }\href@noop {} {\bibfield  {journal} {\bibinfo  {journal} {Computational Materials Science}\ }\textbf {\bibinfo {volume} {199}},\ \bibinfo {pages} {110731} (\bibinfo {year} {2021}{\natexlab{a}})}\BibitemShut {NoStop}%
\bibitem [{\citenamefont {Xie}\ \emph {et~al.}(2021)\citenamefont {Xie}, \citenamefont {Fu}, \citenamefont {Ganea}, \citenamefont {Barzilay},\ and\ \citenamefont {Jaakkola}}]{xie2021crystal}%
  \BibitemOpen
  \bibfield  {author} {\bibinfo {author} {\bibfnamefont {T.}~\bibnamefont {Xie}}, \bibinfo {author} {\bibfnamefont {X.}~\bibnamefont {Fu}}, \bibinfo {author} {\bibfnamefont {O.-E.}\ \bibnamefont {Ganea}}, \bibinfo {author} {\bibfnamefont {R.}~\bibnamefont {Barzilay}}, \ and\ \bibinfo {author} {\bibfnamefont {T.}~\bibnamefont {Jaakkola}},\ }\href@noop {} {\bibfield  {journal} {\bibinfo  {journal} {arXiv preprint arXiv:2110.06197}\ } (\bibinfo {year} {2021})}\BibitemShut {NoStop}%
\bibitem [{\citenamefont {Lyngby}\ and\ \citenamefont {Thygesen}(2022)}]{lyngby2022data}%
  \BibitemOpen
  \bibfield  {author} {\bibinfo {author} {\bibfnamefont {P.}~\bibnamefont {Lyngby}}\ and\ \bibinfo {author} {\bibfnamefont {K.~S.}\ \bibnamefont {Thygesen}},\ }\href@noop {} {\bibfield  {journal} {\bibinfo  {journal} {npj Computational Materials}\ }\textbf {\bibinfo {volume} {8}},\ \bibinfo {pages} {232} (\bibinfo {year} {2022})}\BibitemShut {NoStop}%
\bibitem [{\citenamefont {Wang}\ \emph {et~al.}(2023)\citenamefont {Wang}, \citenamefont {Schmidt}, \citenamefont {Marques}, \citenamefont {Wirtz},\ and\ \citenamefont {Romero}}]{wang2023symmetry}%
  \BibitemOpen
  \bibfield  {author} {\bibinfo {author} {\bibfnamefont {H.-C.}\ \bibnamefont {Wang}}, \bibinfo {author} {\bibfnamefont {J.}~\bibnamefont {Schmidt}}, \bibinfo {author} {\bibfnamefont {M.~A.~L.}\ \bibnamefont {Marques}}, \bibinfo {author} {\bibfnamefont {L.}~\bibnamefont {Wirtz}}, \ and\ \bibinfo {author} {\bibfnamefont {A.~H.}\ \bibnamefont {Romero}},\ }\href {\doibase 10.1088/2053-1583/accc43} {\bibfield  {journal} {\bibinfo  {journal} {2D Materials}\ }\textbf {\bibinfo {volume} {10}},\ \bibinfo {pages} {035007} (\bibinfo {year} {2023})}\BibitemShut {NoStop}%
\bibitem [{\citenamefont {Fu}\ \emph {et~al.}(2024)\citenamefont {Fu}, \citenamefont {Kuisma}, \citenamefont {Larsen}, \citenamefont {Shinohara}, \citenamefont {Togo},\ and\ \citenamefont {Thygesen}}]{fu2024layer}%
  \BibitemOpen
  \bibfield  {author} {\bibinfo {author} {\bibfnamefont {J.}~\bibnamefont {Fu}}, \bibinfo {author} {\bibfnamefont {M.}~\bibnamefont {Kuisma}}, \bibinfo {author} {\bibfnamefont {A.~H.}\ \bibnamefont {Larsen}}, \bibinfo {author} {\bibfnamefont {K.}~\bibnamefont {Shinohara}}, \bibinfo {author} {\bibfnamefont {A.}~\bibnamefont {Togo}}, \ and\ \bibinfo {author} {\bibfnamefont {K.~S.}\ \bibnamefont {Thygesen}},\ }\href@noop {} {\bibfield  {journal} {\bibinfo  {journal} {arXiv preprint arXiv:2401.16705}\ } (\bibinfo {year} {2024})}\BibitemShut {NoStop}%
\bibitem [{\citenamefont {Gjerding}\ \emph {et~al.}(2021{\natexlab{b}})\citenamefont {Gjerding}, \citenamefont {Taghizadeh}, \citenamefont {Rasmussen}, \citenamefont {Ali}, \citenamefont {Bertoldo}, \citenamefont {Deilmann}, \citenamefont {Kn{\o}sgaard}, \citenamefont {Kruse}, \citenamefont {Larsen}, \citenamefont {Manti} \emph {et~al.}}]{gjerding2021recent}%
  \BibitemOpen
  \bibfield  {author} {\bibinfo {author} {\bibfnamefont {M.~N.}\ \bibnamefont {Gjerding}}, \bibinfo {author} {\bibfnamefont {A.}~\bibnamefont {Taghizadeh}}, \bibinfo {author} {\bibfnamefont {A.}~\bibnamefont {Rasmussen}}, \bibinfo {author} {\bibfnamefont {S.}~\bibnamefont {Ali}}, \bibinfo {author} {\bibfnamefont {F.}~\bibnamefont {Bertoldo}}, \bibinfo {author} {\bibfnamefont {T.}~\bibnamefont {Deilmann}}, \bibinfo {author} {\bibfnamefont {N.~R.}\ \bibnamefont {Kn{\o}sgaard}}, \bibinfo {author} {\bibfnamefont {M.}~\bibnamefont {Kruse}}, \bibinfo {author} {\bibfnamefont {A.~H.}\ \bibnamefont {Larsen}}, \bibinfo {author} {\bibfnamefont {S.}~\bibnamefont {Manti}},  \emph {et~al.},\ }\href@noop {} {\bibfield  {journal} {\bibinfo  {journal} {2D Materials}\ }\textbf {\bibinfo {volume} {8}},\ \bibinfo {pages} {044002} (\bibinfo {year} {2021}{\natexlab{b}})}\BibitemShut {NoStop}%
\bibitem [{\citenamefont {Kirklin}\ \emph {et~al.}(2013)\citenamefont {Kirklin}, \citenamefont {Meredig},\ and\ \citenamefont {Wolverton}}]{kirklin2013high}%
  \BibitemOpen
  \bibfield  {author} {\bibinfo {author} {\bibfnamefont {S.}~\bibnamefont {Kirklin}}, \bibinfo {author} {\bibfnamefont {B.}~\bibnamefont {Meredig}}, \ and\ \bibinfo {author} {\bibfnamefont {C.}~\bibnamefont {Wolverton}},\ }\href@noop {} {\bibfield  {journal} {\bibinfo  {journal} {Advanced Energy Materials}\ }\textbf {\bibinfo {volume} {3}},\ \bibinfo {pages} {252} (\bibinfo {year} {2013})}\BibitemShut {NoStop}%
\bibitem [{\citenamefont {Pakdel}\ \emph {et~al.}(2023)\citenamefont {Pakdel}, \citenamefont {Rasmussen}, \citenamefont {Taghizadeh}, \citenamefont {Kruse}, \citenamefont {Olsen},\ and\ \citenamefont {Thygesen}}]{pakdel2023emergent}%
  \BibitemOpen
  \bibfield  {author} {\bibinfo {author} {\bibfnamefont {S.}~\bibnamefont {Pakdel}}, \bibinfo {author} {\bibfnamefont {A.}~\bibnamefont {Rasmussen}}, \bibinfo {author} {\bibfnamefont {A.}~\bibnamefont {Taghizadeh}}, \bibinfo {author} {\bibfnamefont {M.}~\bibnamefont {Kruse}}, \bibinfo {author} {\bibfnamefont {T.}~\bibnamefont {Olsen}}, \ and\ \bibinfo {author} {\bibfnamefont {K.~S.}\ \bibnamefont {Thygesen}},\ }\href@noop {} {\bibfield  {journal} {\bibinfo  {journal} {arXiv preprint arXiv:2304.01148}\ } (\bibinfo {year} {2023})}\BibitemShut {NoStop}%
\bibitem [{\citenamefont {Knobloch}\ \emph {et~al.}(2021)\citenamefont {Knobloch}, \citenamefont {Illarionov}, \citenamefont {Ducry}, \citenamefont {Schleich}, \citenamefont {Wachter}, \citenamefont {Watanabe}, \citenamefont {Taniguchi}, \citenamefont {Mueller}, \citenamefont {Waltl}, \citenamefont {Lanza} \emph {et~al.}}]{knobloch2021performance}%
  \BibitemOpen
  \bibfield  {author} {\bibinfo {author} {\bibfnamefont {T.}~\bibnamefont {Knobloch}}, \bibinfo {author} {\bibfnamefont {Y.~Y.}\ \bibnamefont {Illarionov}}, \bibinfo {author} {\bibfnamefont {F.}~\bibnamefont {Ducry}}, \bibinfo {author} {\bibfnamefont {C.}~\bibnamefont {Schleich}}, \bibinfo {author} {\bibfnamefont {S.}~\bibnamefont {Wachter}}, \bibinfo {author} {\bibfnamefont {K.}~\bibnamefont {Watanabe}}, \bibinfo {author} {\bibfnamefont {T.}~\bibnamefont {Taniguchi}}, \bibinfo {author} {\bibfnamefont {T.}~\bibnamefont {Mueller}}, \bibinfo {author} {\bibfnamefont {M.}~\bibnamefont {Waltl}}, \bibinfo {author} {\bibfnamefont {M.}~\bibnamefont {Lanza}},  \emph {et~al.},\ }\href@noop {} {\bibfield  {journal} {\bibinfo  {journal} {Nature Electronics}\ }\textbf {\bibinfo {volume} {4}},\ \bibinfo {pages} {98} (\bibinfo {year} {2021})}\BibitemShut {NoStop}%
\bibitem [{\citenamefont {Xu}\ \emph {et~al.}(2023)\citenamefont {Xu}, \citenamefont {Liu}, \citenamefont {Liu}, \citenamefont {Zhao}, \citenamefont {Liu}, \citenamefont {Li}, \citenamefont {Nie}, \citenamefont {Liu}, \citenamefont {Yu}, \citenamefont {Feng} \emph {et~al.}}]{xu2023scalable}%
  \BibitemOpen
  \bibfield  {author} {\bibinfo {author} {\bibfnamefont {Y.}~\bibnamefont {Xu}}, \bibinfo {author} {\bibfnamefont {T.}~\bibnamefont {Liu}}, \bibinfo {author} {\bibfnamefont {K.}~\bibnamefont {Liu}}, \bibinfo {author} {\bibfnamefont {Y.}~\bibnamefont {Zhao}}, \bibinfo {author} {\bibfnamefont {L.}~\bibnamefont {Liu}}, \bibinfo {author} {\bibfnamefont {P.}~\bibnamefont {Li}}, \bibinfo {author} {\bibfnamefont {A.}~\bibnamefont {Nie}}, \bibinfo {author} {\bibfnamefont {L.}~\bibnamefont {Liu}}, \bibinfo {author} {\bibfnamefont {J.}~\bibnamefont {Yu}}, \bibinfo {author} {\bibfnamefont {X.}~\bibnamefont {Feng}},  \emph {et~al.},\ }\href@noop {} {\bibfield  {journal} {\bibinfo  {journal} {Nature Materials}\ }\textbf {\bibinfo {volume} {22}},\ \bibinfo {pages} {1078} (\bibinfo {year} {2023})}\BibitemShut {NoStop}%
\bibitem [{\citenamefont {Cao}\ \emph {et~al.}(2019)\citenamefont {Cao}, \citenamefont {Zopf},\ and\ \citenamefont {Ding}}]{Cao_2019}%
  \BibitemOpen
  \bibfield  {author} {\bibinfo {author} {\bibfnamefont {X.}~\bibnamefont {Cao}}, \bibinfo {author} {\bibfnamefont {M.}~\bibnamefont {Zopf}}, \ and\ \bibinfo {author} {\bibfnamefont {F.}~\bibnamefont {Ding}},\ }\href {\doibase 10.1088/1674-4926/40/7/071901} {\bibfield  {journal} {\bibinfo  {journal} {Journal of Semiconductors}\ }\textbf {\bibinfo {volume} {40}},\ \bibinfo {pages} {071901} (\bibinfo {year} {2019})}\BibitemShut {NoStop}%
\bibitem [{\citenamefont {Datta}\ and\ \citenamefont {Das}(1990)}]{datta1990electronic}%
  \BibitemOpen
  \bibfield  {author} {\bibinfo {author} {\bibfnamefont {S.}~\bibnamefont {Datta}}\ and\ \bibinfo {author} {\bibfnamefont {B.}~\bibnamefont {Das}},\ }\href@noop {} {\bibfield  {journal} {\bibinfo  {journal} {Applied Physics Letters}\ }\textbf {\bibinfo {volume} {56}},\ \bibinfo {pages} {665} (\bibinfo {year} {1990})}\BibitemShut {NoStop}%
\bibitem [{\citenamefont {Ashton}\ \emph {et~al.}(2017{\natexlab{b}})\citenamefont {Ashton}, \citenamefont {Gluhovic}, \citenamefont {Sinnott}, \citenamefont {Guo}, \citenamefont {Stewart},\ and\ \citenamefont {Hennig}}]{ashton2017two}%
  \BibitemOpen
  \bibfield  {author} {\bibinfo {author} {\bibfnamefont {M.}~\bibnamefont {Ashton}}, \bibinfo {author} {\bibfnamefont {D.}~\bibnamefont {Gluhovic}}, \bibinfo {author} {\bibfnamefont {S.~B.}\ \bibnamefont {Sinnott}}, \bibinfo {author} {\bibfnamefont {J.}~\bibnamefont {Guo}}, \bibinfo {author} {\bibfnamefont {D.~A.}\ \bibnamefont {Stewart}}, \ and\ \bibinfo {author} {\bibfnamefont {R.~G.}\ \bibnamefont {Hennig}},\ }\href {\doibase 10.1021/acs.nanolett.7b01367} {\bibfield  {journal} {\bibinfo  {journal} {Nano Letters}\ }\textbf {\bibinfo {volume} {17}},\ \bibinfo {pages} {5251} (\bibinfo {year} {2017}{\natexlab{b}})},\ \bibinfo {note} {pMID: 28745061},\ \Eprint {http://arxiv.org/abs/https://doi.org/10.1021/acs.nanolett.7b01367} {https://doi.org/10.1021/acs.nanolett.7b01367} \BibitemShut {NoStop}%
\bibitem [{\citenamefont {Torelli}\ \emph {et~al.}(2019)\citenamefont {Torelli}, \citenamefont {Thygesen},\ and\ \citenamefont {Olsen}}]{torelli2019high}%
  \BibitemOpen
  \bibfield  {author} {\bibinfo {author} {\bibfnamefont {D.}~\bibnamefont {Torelli}}, \bibinfo {author} {\bibfnamefont {K.~S.}\ \bibnamefont {Thygesen}}, \ and\ \bibinfo {author} {\bibfnamefont {T.}~\bibnamefont {Olsen}},\ }\href@noop {} {\bibfield  {journal} {\bibinfo  {journal} {2D Materials}\ }\textbf {\bibinfo {volume} {6}},\ \bibinfo {pages} {045018} (\bibinfo {year} {2019})}\BibitemShut {NoStop}%
\bibitem [{\citenamefont {Tian}\ \emph {et~al.}(2019)\citenamefont {Tian}, \citenamefont {Scullion}, \citenamefont {Hughes}, \citenamefont {Li}, \citenamefont {Shih}, \citenamefont {Coleman}, \citenamefont {Chhowalla},\ and\ \citenamefont {Santos}}]{tian2019electronic}%
  \BibitemOpen
  \bibfield  {author} {\bibinfo {author} {\bibfnamefont {T.}~\bibnamefont {Tian}}, \bibinfo {author} {\bibfnamefont {D.}~\bibnamefont {Scullion}}, \bibinfo {author} {\bibfnamefont {D.}~\bibnamefont {Hughes}}, \bibinfo {author} {\bibfnamefont {L.~H.}\ \bibnamefont {Li}}, \bibinfo {author} {\bibfnamefont {C.-J.}\ \bibnamefont {Shih}}, \bibinfo {author} {\bibfnamefont {J.}~\bibnamefont {Coleman}}, \bibinfo {author} {\bibfnamefont {M.}~\bibnamefont {Chhowalla}}, \ and\ \bibinfo {author} {\bibfnamefont {E.~J.}\ \bibnamefont {Santos}},\ }\href@noop {} {\bibfield  {journal} {\bibinfo  {journal} {Nano Letters}\ }\textbf {\bibinfo {volume} {20}},\ \bibinfo {pages} {841} (\bibinfo {year} {2019})}\BibitemShut {NoStop}%
\bibitem [{\citenamefont {Alvarez}(2013)}]{alvares2013a}%
  \BibitemOpen
  \bibfield  {author} {\bibinfo {author} {\bibfnamefont {S.}~\bibnamefont {Alvarez}},\ }\href {\doibase 10.1039/C3DT50599E} {\bibfield  {journal} {\bibinfo  {journal} {Dalton Trans.}\ }\textbf {\bibinfo {volume} {42}},\ \bibinfo {pages} {8617} (\bibinfo {year} {2013})}\BibitemShut {NoStop}%
\bibitem [{\citenamefont {Osanloo}\ \emph {et~al.}(2022)\citenamefont {Osanloo}, \citenamefont {Saadat}, \citenamefont {Van~de Put}, \citenamefont {Laturia},\ and\ \citenamefont {Vandenberghe}}]{osanloo2022transition}%
  \BibitemOpen
  \bibfield  {author} {\bibinfo {author} {\bibfnamefont {M.~R.}\ \bibnamefont {Osanloo}}, \bibinfo {author} {\bibfnamefont {A.}~\bibnamefont {Saadat}}, \bibinfo {author} {\bibfnamefont {M.~L.}\ \bibnamefont {Van~de Put}}, \bibinfo {author} {\bibfnamefont {A.}~\bibnamefont {Laturia}}, \ and\ \bibinfo {author} {\bibfnamefont {W.~G.}\ \bibnamefont {Vandenberghe}},\ }\href@noop {} {\bibfield  {journal} {\bibinfo  {journal} {Nanoscale}\ }\textbf {\bibinfo {volume} {14}},\ \bibinfo {pages} {157} (\bibinfo {year} {2022})}\BibitemShut {NoStop}%
\bibitem [{\citenamefont {Zhu}\ \emph {et~al.}(2015)\citenamefont {Zhu}, \citenamefont {Wang}, \citenamefont {Xiao}, \citenamefont {Liu}, \citenamefont {Xiong}, \citenamefont {Wong}, \citenamefont {Ye}, \citenamefont {Ye}, \citenamefont {Yin},\ and\ \citenamefont {Zhang}}]{zhu2015observation}%
  \BibitemOpen
  \bibfield  {author} {\bibinfo {author} {\bibfnamefont {H.}~\bibnamefont {Zhu}}, \bibinfo {author} {\bibfnamefont {Y.}~\bibnamefont {Wang}}, \bibinfo {author} {\bibfnamefont {J.}~\bibnamefont {Xiao}}, \bibinfo {author} {\bibfnamefont {M.}~\bibnamefont {Liu}}, \bibinfo {author} {\bibfnamefont {S.}~\bibnamefont {Xiong}}, \bibinfo {author} {\bibfnamefont {Z.~J.}\ \bibnamefont {Wong}}, \bibinfo {author} {\bibfnamefont {Z.}~\bibnamefont {Ye}}, \bibinfo {author} {\bibfnamefont {Y.}~\bibnamefont {Ye}}, \bibinfo {author} {\bibfnamefont {X.}~\bibnamefont {Yin}}, \ and\ \bibinfo {author} {\bibfnamefont {X.}~\bibnamefont {Zhang}},\ }\href@noop {} {\bibfield  {journal} {\bibinfo  {journal} {Nature Nanotechnology}\ }\textbf {\bibinfo {volume} {10}},\ \bibinfo {pages} {151} (\bibinfo {year} {2015})}\BibitemShut {NoStop}%
\bibitem [{\citenamefont {Wu}\ \emph {et~al.}(2014)\citenamefont {Wu}, \citenamefont {Wang}, \citenamefont {Li}, \citenamefont {Zhang}, \citenamefont {Lin}, \citenamefont {Niu}, \citenamefont {Chenet}, \citenamefont {Zhang}, \citenamefont {Hao}, \citenamefont {Heinz} \emph {et~al.}}]{wu2014piezoelectricity}%
  \BibitemOpen
  \bibfield  {author} {\bibinfo {author} {\bibfnamefont {W.}~\bibnamefont {Wu}}, \bibinfo {author} {\bibfnamefont {L.}~\bibnamefont {Wang}}, \bibinfo {author} {\bibfnamefont {Y.}~\bibnamefont {Li}}, \bibinfo {author} {\bibfnamefont {F.}~\bibnamefont {Zhang}}, \bibinfo {author} {\bibfnamefont {L.}~\bibnamefont {Lin}}, \bibinfo {author} {\bibfnamefont {S.}~\bibnamefont {Niu}}, \bibinfo {author} {\bibfnamefont {D.}~\bibnamefont {Chenet}}, \bibinfo {author} {\bibfnamefont {X.}~\bibnamefont {Zhang}}, \bibinfo {author} {\bibfnamefont {Y.}~\bibnamefont {Hao}}, \bibinfo {author} {\bibfnamefont {T.~F.}\ \bibnamefont {Heinz}},  \emph {et~al.},\ }\href@noop {} {\bibfield  {journal} {\bibinfo  {journal} {Nature}\ }\textbf {\bibinfo {volume} {514}},\ \bibinfo {pages} {470} (\bibinfo {year} {2014})}\BibitemShut {NoStop}%
\bibitem [{\citenamefont {Olsen}\ \emph {et~al.}(2019)\citenamefont {Olsen}, \citenamefont {Andersen}, \citenamefont {Okugawa}, \citenamefont {Torelli}, \citenamefont {Deilmann},\ and\ \citenamefont {Thygesen}}]{olsen2019discovering}%
  \BibitemOpen
  \bibfield  {author} {\bibinfo {author} {\bibfnamefont {T.}~\bibnamefont {Olsen}}, \bibinfo {author} {\bibfnamefont {E.}~\bibnamefont {Andersen}}, \bibinfo {author} {\bibfnamefont {T.}~\bibnamefont {Okugawa}}, \bibinfo {author} {\bibfnamefont {D.}~\bibnamefont {Torelli}}, \bibinfo {author} {\bibfnamefont {T.}~\bibnamefont {Deilmann}}, \ and\ \bibinfo {author} {\bibfnamefont {K.~S.}\ \bibnamefont {Thygesen}},\ }\href {\doibase 10.1103/PhysRevMaterials.3.024005} {\bibfield  {journal} {\bibinfo  {journal} {Phys. Rev. Mater.}\ }\textbf {\bibinfo {volume} {3}},\ \bibinfo {pages} {024005} (\bibinfo {year} {2019})}\BibitemShut {NoStop}%
\bibitem [{\citenamefont {Castro~Neto}\ \emph {et~al.}(2009)\citenamefont {Castro~Neto}, \citenamefont {Guinea}, \citenamefont {Peres}, \citenamefont {Novoselov},\ and\ \citenamefont {Geim}}]{neto2009the}%
  \BibitemOpen
  \bibfield  {author} {\bibinfo {author} {\bibfnamefont {A.~H.}\ \bibnamefont {Castro~Neto}}, \bibinfo {author} {\bibfnamefont {F.}~\bibnamefont {Guinea}}, \bibinfo {author} {\bibfnamefont {N.~M.~R.}\ \bibnamefont {Peres}}, \bibinfo {author} {\bibfnamefont {K.~S.}\ \bibnamefont {Novoselov}}, \ and\ \bibinfo {author} {\bibfnamefont {A.~K.}\ \bibnamefont {Geim}},\ }\href {\doibase 10.1103/RevModPhys.81.109} {\bibfield  {journal} {\bibinfo  {journal} {Rev. Mod. Phys.}\ }\textbf {\bibinfo {volume} {81}},\ \bibinfo {pages} {109} (\bibinfo {year} {2009})}\BibitemShut {NoStop}%
\bibitem [{\citenamefont {Kharadi}\ \emph {et~al.}(2020)\citenamefont {Kharadi}, \citenamefont {Malik}, \citenamefont {Khanday}, \citenamefont {Shah}, \citenamefont {Mittal},\ and\ \citenamefont {Kaushik}}]{kharadi2020review}%
  \BibitemOpen
  \bibfield  {author} {\bibinfo {author} {\bibfnamefont {M.~A.}\ \bibnamefont {Kharadi}}, \bibinfo {author} {\bibfnamefont {G.~F.~A.}\ \bibnamefont {Malik}}, \bibinfo {author} {\bibfnamefont {F.~A.}\ \bibnamefont {Khanday}}, \bibinfo {author} {\bibfnamefont {K.~A.}\ \bibnamefont {Shah}}, \bibinfo {author} {\bibfnamefont {S.}~\bibnamefont {Mittal}}, \ and\ \bibinfo {author} {\bibfnamefont {B.~K.}\ \bibnamefont {Kaushik}},\ }\href {\doibase 10.1149/2162-8777/abd09a} {\bibfield  {journal} {\bibinfo  {journal} {ECS Journal of Solid State Science and Technology}\ }\textbf {\bibinfo {volume} {9}},\ \bibinfo {pages} {115031} (\bibinfo {year} {2020})}\BibitemShut {NoStop}%
\bibitem [{\citenamefont {Liu}\ \emph {et~al.}(2011)\citenamefont {Liu}, \citenamefont {Feng},\ and\ \citenamefont {Yao}}]{liu2011quantum}%
  \BibitemOpen
  \bibfield  {author} {\bibinfo {author} {\bibfnamefont {C.-C.}\ \bibnamefont {Liu}}, \bibinfo {author} {\bibfnamefont {W.}~\bibnamefont {Feng}}, \ and\ \bibinfo {author} {\bibfnamefont {Y.}~\bibnamefont {Yao}},\ }\href {\doibase 10.1103/PhysRevLett.107.076802} {\bibfield  {journal} {\bibinfo  {journal} {Phys. Rev. Lett.}\ }\textbf {\bibinfo {volume} {107}},\ \bibinfo {pages} {076802} (\bibinfo {year} {2011})}\BibitemShut {NoStop}%
\bibitem [{\citenamefont {Mortensen}\ \emph {et~al.}(2020)\citenamefont {Mortensen}, \citenamefont {Gjerding},\ and\ \citenamefont {Thygesen}}]{mortensen2020myqueue}%
  \BibitemOpen
  \bibfield  {author} {\bibinfo {author} {\bibfnamefont {J.}~\bibnamefont {Mortensen}}, \bibinfo {author} {\bibfnamefont {M.}~\bibnamefont {Gjerding}}, \ and\ \bibinfo {author} {\bibfnamefont {K.}~\bibnamefont {Thygesen}},\ }\href@noop {} {\bibfield  {journal} {\bibinfo  {journal} {The Journal of Open Source Software}\ }\textbf {\bibinfo {volume} {5}},\ \bibinfo {pages} {1844} (\bibinfo {year} {2020})}\BibitemShut {NoStop}%
\bibitem [{\citenamefont {Enkovaara}\ \emph {et~al.}(2010)\citenamefont {Enkovaara}, \citenamefont {Rostgaard}, \citenamefont {Mortensen}, \citenamefont {Chen}, \citenamefont {Du{\l}ak}, \citenamefont {Ferrighi}, \citenamefont {Gavnholt}, \citenamefont {Glinsvad}, \citenamefont {Haikola}, \citenamefont {Hansen} \emph {et~al.}}]{enkovaara2010electronic}%
  \BibitemOpen
  \bibfield  {author} {\bibinfo {author} {\bibfnamefont {J.}~\bibnamefont {Enkovaara}}, \bibinfo {author} {\bibfnamefont {C.}~\bibnamefont {Rostgaard}}, \bibinfo {author} {\bibfnamefont {J.~J.}\ \bibnamefont {Mortensen}}, \bibinfo {author} {\bibfnamefont {J.}~\bibnamefont {Chen}}, \bibinfo {author} {\bibfnamefont {M.}~\bibnamefont {Du{\l}ak}}, \bibinfo {author} {\bibfnamefont {L.}~\bibnamefont {Ferrighi}}, \bibinfo {author} {\bibfnamefont {J.}~\bibnamefont {Gavnholt}}, \bibinfo {author} {\bibfnamefont {C.}~\bibnamefont {Glinsvad}}, \bibinfo {author} {\bibfnamefont {V.}~\bibnamefont {Haikola}}, \bibinfo {author} {\bibfnamefont {H.}~\bibnamefont {Hansen}},  \emph {et~al.},\ }\href@noop {} {\bibfield  {journal} {\bibinfo  {journal} {Journal of Physics: Condensed Matter}\ }\textbf {\bibinfo {volume} {22}},\ \bibinfo {pages} {253202} (\bibinfo {year} {2010})}\BibitemShut {NoStop}%
\bibitem [{\citenamefont {Larsen}\ \emph {et~al.}(2017)\citenamefont {Larsen}, \citenamefont {Mortensen}, \citenamefont {Blomqvist}, \citenamefont {Castelli}, \citenamefont {Christensen}, \citenamefont {Du{\l}ak}, \citenamefont {Friis}, \citenamefont {Groves}, \citenamefont {Hammer}, \citenamefont {Hargus} \emph {et~al.}}]{larsen2017atomic}%
  \BibitemOpen
  \bibfield  {author} {\bibinfo {author} {\bibfnamefont {A.~H.}\ \bibnamefont {Larsen}}, \bibinfo {author} {\bibfnamefont {J.~J.}\ \bibnamefont {Mortensen}}, \bibinfo {author} {\bibfnamefont {J.}~\bibnamefont {Blomqvist}}, \bibinfo {author} {\bibfnamefont {I.~E.}\ \bibnamefont {Castelli}}, \bibinfo {author} {\bibfnamefont {R.}~\bibnamefont {Christensen}}, \bibinfo {author} {\bibfnamefont {M.}~\bibnamefont {Du{\l}ak}}, \bibinfo {author} {\bibfnamefont {J.}~\bibnamefont {Friis}}, \bibinfo {author} {\bibfnamefont {M.~N.}\ \bibnamefont {Groves}}, \bibinfo {author} {\bibfnamefont {B.}~\bibnamefont {Hammer}}, \bibinfo {author} {\bibfnamefont {C.}~\bibnamefont {Hargus}},  \emph {et~al.},\ }\href@noop {} {\bibfield  {journal} {\bibinfo  {journal} {Journal of Physics: Condensed Matter}\ }\textbf {\bibinfo {volume} {29}},\ \bibinfo {pages} {273002} (\bibinfo {year} {2017})}\BibitemShut {NoStop}%
\bibitem [{\citenamefont {Perdew}\ \emph {et~al.}(1996)\citenamefont {Perdew}, \citenamefont {Burke},\ and\ \citenamefont {Ernzerhof}}]{perdew1996generalized}%
  \BibitemOpen
  \bibfield  {author} {\bibinfo {author} {\bibfnamefont {J.~P.}\ \bibnamefont {Perdew}}, \bibinfo {author} {\bibfnamefont {K.}~\bibnamefont {Burke}}, \ and\ \bibinfo {author} {\bibfnamefont {M.}~\bibnamefont {Ernzerhof}},\ }\href@noop {} {\bibfield  {journal} {\bibinfo  {journal} {Physical review letters}\ }\textbf {\bibinfo {volume} {77}},\ \bibinfo {pages} {3865} (\bibinfo {year} {1996})}\BibitemShut {NoStop}%
\bibitem [{\citenamefont {Heyd}\ \emph {et~al.}(2003)\citenamefont {Heyd}, \citenamefont {Scuseria},\ and\ \citenamefont {Ernzerhof}}]{heyd2003hybrid}%
  \BibitemOpen
  \bibfield  {author} {\bibinfo {author} {\bibfnamefont {J.}~\bibnamefont {Heyd}}, \bibinfo {author} {\bibfnamefont {G.~E.}\ \bibnamefont {Scuseria}}, \ and\ \bibinfo {author} {\bibfnamefont {M.}~\bibnamefont {Ernzerhof}},\ }\href@noop {} {\bibfield  {journal} {\bibinfo  {journal} {The Journal of Chemical Physics}\ }\textbf {\bibinfo {volume} {118}},\ \bibinfo {pages} {8207} (\bibinfo {year} {2003})}\BibitemShut {NoStop}%
\end{thebibliography}%

\end{document}